\documentclass[12pt,letter]{article}

\usepackage{graphicx, epsfig, color}
\def\eslt{\not\!\!{E_T}}
\def\to{\rightarrow}

\def\bi{\begin{itemize}}
\def\ei{\end{itemize}}
\def\te{\tilde e}

\def\tu{\tilde u}
\def\sps1ap{SPS1a$^\prime$}
\def\c1p{C1$^\prime$}

\def\cO{{\cal O}}

\def\tb{\tilde b}

\def\td{\tilde d}

\def\tst{\tilde t}
\def\ttau{\tilde \tau}

\def\tg{\tilde g}

\def\tell{\tilde\ell}
\def\tq{\tilde q}
\def\pino{\tilde\gamma}
\def\tw{\widetilde W}
\def\tz{\widetilde Z}
\def\alt{\stackrel{<}{\sim}}
\def\agt{\stackrel{>}{\sim}}
\def\be{\begin{equation}}  
\def\ee{\end{equation}}  
\def\bea{\begin{eqnarray}}  
\def\eea{\end{eqnarray}}  
\def\beas{\begin{eqnarray*}}  
\def\eeas{\end{eqnarray*}}  
\newcommand\prd[3]{{\it Phys.\ Rev.\ }{\bf D #1} (#2) #3}

\newcommand\prl[3]{{\it Phys.\ Rev.\ Lett.\ }{\bf #1} (#2) #3}
\newcommand\plb[3]{{\it Phys.\ Lett.\ }{\bf B #1} (#2) #3}
\newcommand\jhep[3]{{\it J. High Energy Phys.\ }{\bf #1} (#2) #3}

\newcommand\npb[3]{{\it Nucl.\ Phys.\ }{\bf B #1} (#2) #3}
\newcommand\epjc[3]{{\it Eur.\ Phys.\ J. }{\bf C #1} (#2) #3}

\newcommand{\hepph}[1]{hep-ph/#1}

\newcommand\ppnp[3]{{\it Prog.\ Part.\ Nucl.\ Phys.}{\bf  #1} (#2) #3}

\begin{document}
\begin{titlepage}

\vspace{0.5cm}
\begin{center}
{\Large \bf Upper bounds on sparticle masses from naturalness\\
or how to disprove weak scale supersymmetry
}\\ 
\vspace{1.2cm} \renewcommand{\thefootnote}{\fnsymbol{footnote}}
{\large Howard Baer$^1$\footnote[1]{Email: baer@nhn.ou.edu }, 
Vernon Barger$^2$\footnote[2]{Email: barger@pheno.wisc.edu },
and Michael Savoy$^1$\footnote[3]{Email: savoy@nhn.ou.edu }
}\\ 
\vspace{1.2cm} \renewcommand{\thefootnote}{\arabic{footnote}}
{\it 
$^1$Dept. of Physics and Astronomy,
University of Oklahoma, Norman, OK 73019, USA \\
}
{\it 
$^2$Dept. of Physics,
University of Wisconsin, Madison, WI 53706, USA \\
}

\end{center}

\vspace{0.5cm}
\begin{abstract}
While it is often stated that the notion of electroweak (EW) naturalness 
in supersymmetric models is subjective, fuzzy and model-dependent,
here we argue the contrary: electroweak naturalness can be elevated to a 
{\it principle} which is both objective and predictive.
We demonstrate visually when too much fine-tuning sets in at the electroweak scale
which corresponds numerically to the measure $\Delta_{BG}\sim \Delta_{EW}\agt 30$.
While many constrained SUSY models are already excluded by this value, 
we derive updated upper bounds on sparticle masses within the
two-extra parameter non-universal Higgs model (NUHM2). 
We confirm the classic Barbieri-Giudice (BG) result that $\Delta_{BG}<30$ implies $\mu <350$ GeV.
However, by combining dependent soft terms which appear as multiples of $m_{3/2}$ in supergravity
models, then we obtain $m_{\tg}\alt 4$ TeV as opposed to the BG result that
$m_{\tg}\alt 350$ GeV.
We compare the NUHM2 results to a similar scan in the pMSSM with 19 weak scale parameters. 
In the pMSSM with complete one-loop scalar potential plus dominant two-loop terms, 
then a $m_{\tg}<7$ TeV bound is found.
Our tabulation of upper bounds provides a target for experimenters seeking to discover or else falsify 
the existence of weak scale supersymmetry.
In an Appendix, we show contributions to the naturalness measure from one-loop
contributions to the weak scale  scalar potential.
\noindent 
\vspace*{0.8cm}

\end{abstract}

\end{titlepage}

\section{Introduction}
\label{sec:intro}

Weak scale supersymmetry\cite{wss} (SUSY) is a highly motivated paradigm for physics beyond the Standard Model (SM).
The principal motivation is that it offers a technical solution to the gauge hierarchy problem or
naturalness problem of the SM in that it ensures cancellation of quadratic divergences of 
scalar field masses to all orders in perturbation theory\cite{kaul}. 
This is especially relevant now that a
bonafide fundamental scalar has been discovered\cite{atlas_h,cms_h}: 
{\it i.e.} the Higgs boson $h$.
The SUSY technical solution is accomplished in a highly simple manner: merely extending the 
set of spacetime symmetries which underlies quantum field theory to their most general structure
based upon a graded Lie algebra. 
In fact the SUSY paradigm is supported indirectly via three sets of measurements:
\begin{enumerate}
\item the measured values of the three gauge forces at the weak scale are exactly what is needed 
for SUSY gauge coupling unification\cite{gauge},
\item the measured value of the top quark mass is exactly what is needed to properly 
drive a radiative breakdown in electroweak symmetry\cite{rewsb} and
\item the measured value of the Higgs boson mass $m_h=125.09\pm 0.24$ GeV falls squarely within the 
predicted narrow band of allowed MSSM values\cite{mhiggs}.
\end{enumerate}

On the negative side of the ledger, there is as of yet no sign of supersymmetric matter after extensive runs
of LHC at $\sqrt{s}=7-8$ TeV\cite{atlas_s,cms_s}. 
This fact has led many theorists to question the validity of SUSY in light of 
the oft-repeated mantra that weak scale naturalness requires weak scale sparticles. 
It is also baffling on the experimental side as to when the weak scale SUSY hypothesis is ruled out, 
and when one ought to move on to alternative directions. 
This is especially vexing in that many  theoretical
predictions tend to lie just beyond current exclusion limits: when the exclusion limits increase, then the
theoretical predictions retreat towards higher mass values again just beyond the latest lower mass bounds.
This leads to the important question: {\it what does it take to falsify the weak scale SUSY hypothesis}? 
When is the job done? Are new accelerators and experiments required, 
or does LHC with $\sqrt{s}\sim 13-14$ TeV and high luminosity have the necessary resolving power?

Historically, an answer to this question was provided in the classic paper by 
Barbieri and Giudice\cite{bg} wherein-- based upon the ``naturalness criterion''-- 
they derived upper bounds on various sparticle masses.\footnote{See also Ref. \cite{AC}.} 
These upper bounds could serve as targets for experimental
facilities with the intent to discover or disprove weak scale SUSY. 
Following earlier work by Ellis {\it et al.}\cite{eenz}, they introduced the naturalness measure\footnote{The authors of Ref's \cite{bayes} note the link between naturalness and Bayesian 
statistics wherein naturalness corresponds to ``more probable''.}
\be
\Delta_{BG}\equiv max_i \left| \frac{\partial\log m_Z^2}{\partial\log p_i} \right|
\ee
where the $p_i$ are fundamental parameters of the theory labelled by the index $i$. 
Working within the MSSM with unified GUT scale soft breaking terms $m_0,\ m_{1/2}, A_0$ and $B$, 
they presented upper bounds on sparticle masses and parameters as a function of the top-quark mass 
(which of course was not yet known at the time). 
In light of our present knowledge of the top quark mass 
($m_t=173.21\pm 0.87$ GeV PDG value with combined statistical/systematic errors), they would conclude
that in order to accommodate $\Delta_{BG}<10$ (or better than 10\% fine-tuning in $m_Z$), then
$m_0\alt 300$ GeV, $m_{1/2}\alt 100$ GeV and $\mu\alt 200$ GeV. These upper bounds implied
that the lighter charginos $m_{\tw_1}\alt 100$ GeV and gluinos $m_{\tg}\alt 350$ GeV. 
Explorations at LEP2 resulted in limits of $m_{\tw_1}>103.5$ GeV\cite{lep2ino} 
so that weak scale SUSY already appeared somewhat fine-tuned by the post-LEP and pre-LHC era\cite{LH}. 
The current limits on gluino mass from LHC8 require $m_{\tg}\agt 1300$ GeV. 
Thus, in the post LHC8 era, the naturalness issue has intensified, and based 
upon these theory/experiment confrontations, 
one might well be tempted to conclude that weak scale SUSY has been disproved\cite{lykken}.

In this paper, we re-examine upper bounds on sparticle masses from the naturalness principle. 
In Sec. \ref{sec:nat}, we discuss naturalness and fine-tuning and articulate the Naturalness Principle.
In Sec. \ref{sec:dew}, we apply the Naturalness Principle at the weak scale to derive the
electroweak fine-tuning measure $\Delta_{EW}$. In Sec. \ref{sec:BG} we discuss Barbieri-Giudice
fine-tuning $\Delta_{BG}$, how it depends on the selection of an independent parameter set 
 and how it relates to $\Delta_{EW}$: when properly applied by combining dependent contributions, 
then $\Delta_{BG}\simeq \Delta_{EW}$. 
In Sec. \ref{sec:30}, we demonstrate visually when too much fine-tuning sets in and what values
of $\Delta_{EW}\simeq \Delta_{BG}$ are too much.
In Sec. \ref{sec:mass} we derive upper bounds on sparticle masses in 
the NUHM2 model and compare them to upper bounds from the classic BG paper. In Sec. \ref{sec:pmssm},
we derive alternative upper bounds arising from a scan over the 19 dimensional weak scale pMSSM 
parameter space. We compare these against results from the 19 parameter SUGRA model
with soft terms defined at $Q=m_{GUT}$.
In Sec. \ref{sec:conclude}, we conclude by presenting a bar chart of upper bounds on sparticle masses
which form a target for experimenters seeking to confirm or refute weak scale supersymmetry.
The hard target is that $\mu <350$ GeV in any models (defined at a high scale or weak scale) 
based on the MSSM. In models with
RG running from $m_{GUT}$ to $m_{weak}$, then $m_{\tg}\alt 2$ (4) TeV for 
$\Delta_{BG}\simeq\Delta_{EW}<10$ (30). If one dispenses with RG running as in the pMSSM, 
then no bound is obtained on $m_{\tg}$ using the one-loop RG-improved effective potential.
By including leading two-loop contrubutions to the scalar potential, we find a bound
of $m_{\tg}<7$ TeV for $\Delta_{EW}<30$ in the pMSSM.
In an Appendix, we discuss the contributions to $\Delta_{EW}$ from various 
radiative corrections $\Sigma_u^u(i)$ 
and plot their magnitudes in 2-dimensional parameter planes.

\section{The naturalness principle}
\label{sec:nat}

Several definitions of naturalness can be found in the literature: 
some are more abstract while others are more pragmatic. 
Here we articulate the following Naturalness Principle\footnote{
Dimopoulos and Susskind articulate:``Naturalness: no parameter needs to be adjusted to unreasonable accuracy.''\cite{dim_suss}}:\footnote{
Weinberg states: ``The appearance of fine-tuning in a scientific theory 
is like a cry of distress from nature, complaining that something needs to be better explained''\cite{sweinberg}.}
\begin{quote}
\item An observable $\cO$ is {\bf natural} if all {\it independent} contributions to ${\cal O}$
are less than or of order ${\cal O}$.
\end{quote}

Suppose $\cO$ can be calculated in terms of $n$ independent contributions $\cO = o_{1} +\cdots +o_{n}$.
If one of the contributions, say $o_n$, is far larger than $\cO$, then it would have to be the case
that one or more of the $o_i$ would have to be a large opposite sign contribution which would require
fine-tuning of $o_i\sim -o_n$ such that $\cO\ll o_n$. This is the link between naturalness and fine-tuning:
a quantity $\cO$ is natural if it requires no large fine-tuning of independent contributions to maintain
its measured value. 

A common pitfall in evaluating when a quantity is natural is to split it into {\it dependent} contributions:
if $\cO=o_1+o_2+\cdots$, but as one increases $o_1$ then $o_2$ necessarily increases to large opposite-sign, then
$o_1$ and $o_2$ should be combined before the final evaluation of naturalness. This pitfall has been dubbed the 
``fine-tuning rule''\cite{seige}.

To see how $\Delta_{BG}$ is a measure of naturalness, consider the observable $\cO$ expressed as a linear
combination of $n$ fundamental parameters $p_i$: 
\be
\cO= a_1p_1+\cdots +a_n p_n
\label{eq:aipi}
\ee 
where the $a_i$ are numerical co-efficients. 
Applying the $\Delta_{BG}$ measure, we would find
\be
\Delta_{BG}=max_i\left|\frac{p_i}{\cO}\frac{\partial \cO}{\partial p_i}\right|=max_i\left| a_ip_i/\cO \right|
\ee
{\it i.e.} the BG measure just picks off each term on the right-hand-side of Eq. \ref{eq:aipi}
and compares it to $\cO$.\footnote{If terms include powers of $p_i$, {\it e.g.} $\cO=\sum_i a_ip_i^{n_i}$, 
then each sensitivity co-efficient contains an additional factor of $n_i$.} 
To avoid fine-tunings, then $\Delta_{BG}$ should be less than some
value (typically 10-100) depending on how much fine-tuning one is willing to tolerate.

\section{Weak scale naturalness}
\label{sec:dew}

Starting with the weak scale scalar (Higgs) potential of the MSSM, the minimization conditions
$\frac{\partial V}{\partial h_u^{0*}}=\frac{\partial V}{\partial h_d^{0*}}=0$ allow one to determine the 
Higgs field VEVs in terms of the soft SUSY breaking parameters and the $\mu$ parameter\cite{wss}. 
Then, since 
$m_Z^2=(g^2+g^{\prime 2})(v_u^2+v_d^2)/2$, we can relate the observed value of $m_Z$ to the 
weak scale SUSY parameters as
\be
\frac{m_Z^2}{2} = \frac{(m_{H_d}^2+\Sigma_d^d)-(m_{H_u}^2+\Sigma_u^u)\tan^2\beta}{(\tan^2\beta -1)}
-\mu^2\simeq -m_{H_u}^2-\mu^2 .
\label{eq:mzs}
\ee
Here, $m_{H_u}^2$ and $m_{H_d}^2$ are the {\it weak scale} soft SUSY breaking Higgs masses, $\mu$ 
is the {\it supersymmetric} higgsino mass term and $\Sigma_u^u$ and $\Sigma_d^d$ contain
an assortment of loop corrections to the effective potential.\footnote{We will not consider $\tan\beta$
as an independent parameter here. Its value is determined also by the minimization conditions in terms of
the SUSY parameters.} 
Already at this stage (not yet worrying about high scale parameters),
the naturalness principle requires each term on the right-hand-side (RHS) 
of Eq. \ref{eq:mzs} to be comparable to or less 
than $m_Z^2/2$. 
From the partial equality on the RHS, it is plain to see that the weak scale values of
$m_{H_u}^2$, $\mu^2$ and the various $\Sigma_u^u(i)$ ($i$ labels the various loop contributions) 
should all be comparable to or less than $m_Z^2/2$\cite{ccn}. This allows us to define the
{\it electroweak} fine-tuning measure $\Delta_{EW}$ as
\be
\Delta_{EW}=max\left| each\ term\ on\ RHS\ of\ Eq.~\ref{eq:mzs}\right| 
/(m_Z^2/2) .
\label{eq:dew}
\ee 

In gravity-mediation, the GUT scale soft terms are expected to be of order the gravitino mass $m_{3/2}$. 
Since the physical sparticle masses are derived from the soft terms, we thus expect the weak scale soft terms 
also to be of order $m_{3/2}$. 
Since LHC8 requires multi-TeV values of $m_{\tg}$ and $m_{\tq}$, then it seems LHC8 is telling
us that $m_{3/2}$ is also in the multi-TeV range. This is good news for the SUSY flavor and 
CP problems\cite{dine} and the gravitino problem\cite{grav}: 
all these are ameliorated by multi-TeV values of $m_{3/2}$ and sparticle masses.
The puzzle then is: why is $m_Z$ (and $m_W$ and $m_h$) not also at the multi-TeV scale? 
There is one soft term whose weak scale value may be very different from its GUT scale value. 
The term $m_{H_u}^2$ may hold multi-TeV values at $m_{GUT}$ but is necessarily driven 
through zero to negative values at $Q= m_{weak}$ in order to break electroweak symmetry\cite{rewsb}. 
If it is driven to small values comparable to $-m_Z^2/2$ 
rather than large negative values, then one naturalness condition may be satisfied.
This case has been dubbed {\it radiatively-driven naturalness}\cite{ltr,rns}.

The other condition for naturalness is that $\mu^2\sim m_Z^2/2$.\footnote{In this regard, we rely
on Einstein's advice to maintain the theory as simple as possible, but no simpler.
By adding various exotica to the MSSM, then one can create models with heavy higgsinos
which may still be natural\cite{nelson,luty,spmartin}. These often involve adding exotic states such as
scalar gluons. It is not clear whether such constructs admit a UV-completion.} 
In this case, we note that $\mu$ is SUSY conserving and not SUSY breaking. 
It is a very different entity from the soft terms: naively, it would be present
even in the absence of SUSY breaking. In this case, 
one expects its value to be of order the GUT or Planck scale $M_P$.
How it comes instead to be $\sim m_Z$ is known as the SUSY $\mu$ problem\cite{kn}. 
There are two parts to the solution of the SUSY $\mu$ problem: first, one must forbid it (via some symmetry) 
from attaining values $\sim M_P$, and second, one must regenerate it
of order $m_Z$ (via symmetry breaking). In the original Kim-Nilles formulation\cite{kn}, 
it was noted that in the SUSY DFSZ\cite{dfsz} axion model, 
Peccei-Quinn symmetry forbids the $\mu$ term. 
The spontaneous breaking of PQ symmetry at scale $f_a\sim 10^{11}$ GeV generates an axion 
(thus solving the notorious strong CP problem), 
but also generates a mu term of order $f_a^2/M_P$\cite{chun}. 
Since one expects the gravitino mass (and hence soft terms) to be
of order $m_{3/2}\sim m_{hidden}^2/ M_P$, where $m_{hidden}$ is an intermediate mass scale associated 
with hidden sector SUGRA breaking, 
then the apparent Little Hierarchy $\mu\ll m_{3/2}$ is just a consequence of a mismatch
between PQ breaking scale and hidden sector mass scale: $f_a\ll m_{hidden}$. 
In fact, in models such as the Murayama-Suzuki-Yanagida (MSY) SUSY axion model\cite{msy}, 
PQ symmetry is radiatively broken as a consequence of SUSY breaking.
In the MSY model, for canonical parameter values a small value of $\mu\sim 100-200$ GeV can be 
easily generated from $m_{3/2}\sim 5-20$ TeV\cite{radpq}.

Finally, the radiative corrections $\Sigma_u^u$ should be less than or $\sim m_Z^2/2$. Typically, the 
top squark contributions are the largest of these. 
The top squark contributions $\Sigma_u^u(\tst_{1,2})$ are minimized for TeV-scale highly
mixed top squarks, which also lift the Higgs mass to $m_h\sim 125$ GeV~\cite{ltr}.

The EW fine-tuning measure is the most conservative of the fine-tuning measures in that
any model with large $\Delta_{EW}$ is surely fine-tuned. Also, $\Delta_{EW}$ clearly agrees 
with $\Delta_{BG}$ for SUSY models defined purely at the weak scale such as the 
pMSSM.\cite{pmssm}\footnote{The pMSSM, or phenomenological MSSM, is the MSSM
defined with weak scale input parameters where all CP violating and flavor violating soft terms have
been set to zero. Also, usually first/second generation soft terms are set equal
to each other to avoid flavor-violations.}. 
Another virtue of $\Delta_{EW}$ is that its value is {\it model-independent} (within the MSSM) in that
it doesn't depend on {\it how} the weak scale spectrum was generated. 
Also, from a pragmatic point of view, the fine-tuning encapsulated in $\Delta_{EW}$ is exactly where
spectrum generators invoke explicit fine-tuning. In such codes, usually $m_{H_u}^2$ and other soft terms are
calculated at the weak scale via RG running, and then the value of $\mu$ is dialed/fine-tuned to enforce that
the measured value of $m_Z$ is obtained. Without such fine-tuning, the generated value of $m_Z$ would typically
lie in the multi-TeV region\cite{seige}.

But does $\Delta_{EW}$ encapsulate {\it all} the fine-tuning, including high-scale effects, 
or is it just a lower bound on fine-tuning\cite{mt}? 

\section{BG fine-tuning and independent model parameters}
\label{sec:BG}

For models defined in terms of high scale parameters, the BG measure can be evaluated by 
expanding the terms on the RHS of Eq. \ref{eq:mzs}
using semi-analytic RG solutions in terms of fundamental high scale parameters~\cite{bg,munoz}.
For the case of $\tan\beta =10$ and taking the high scale $\Lambda =m_{GUT}$, then one finds~\cite{abe,martin,feng}
\be
m_Z^2 \simeq  -2.18\mu^2 + 3.84 M_3^2-0.65 M_3A_t-1.27 m_{H_u}^2 -0.053 m_{H_d}^2
+0.73 m_{Q_3}^2+0.57 m_{U_3}^2 +\cdots
\label{eq:mzsHS}
\ee

The problem with most applications of the BG measure  is that in any sensible model of SUSY breaking, 
the high scale SUSY parameters are not independent. For instance in gravity-mediation,
for any given hidden sector, the soft SUSY breaking terms are all calculated as numerical
co-efficients times the gravitino mass~\cite{sw,kl,Brignole:1993dj}: {\it e.g.} $M_3(\Lambda )=a_{M_3} m_{3/2}$, 
$A_t=a_{A_t}m_{3/2}$, $m_{Q_3}^2=a_{Q_3}m_{3/2}^2$, {\it etc.} where the $a_i$ are just numerical constants. 
(For example, in string theory with dilaton-dominated SUSY breaking~\cite{kl,Brignole:1993dj}, 
then we expect $m_0^2=m_{3/2}^2$ with $m_{1/2}=-A_0=\sqrt{3}m_{3/2}$).
The reason one scans multiple SUSY model soft term parameters is to account for a 
wide variety of hidden sectors possibilities. 
But this doesn't mean each soft term is independent from the others.
By writing the soft terms in Eq. \ref{eq:mzsHS} as suitable multiples of $m_{3/2}^2$, 
then large positive and negative contributions can be combined/cancelled 
and one arrives at the simpler expression~\cite{seige,arno}:
\be
m_Z^2\simeq -2\mu^2(\Lambda) +a\cdot m_{3/2}^2 
\label{eq:mzssugra}
\ee
since the SUSY mu term hardly evolves.
The value of $a$ is just some number which is the sum of all the coefficients of the terms 
$\propto m_{3/2}^2$.

Using the BG measure applied to Eq. \ref{eq:mzssugra}, then it is found that naturalness requires 
$\mu^2\sim \mu^2(\Lambda )\sim m_Z^2$ and also that $am_{3/2}^2\sim m_Z^2$. The first requirement is the same as in
$\Delta_{EW}$. The second requirement is fulfilled {\it either} by $m_{3/2}\sim m_Z$ 
(which seems unlikely in light of LHC Higgs mass measurement and sparticle mass bounds) {\it or}
when $m_{3/2}$ is large but the co-efficient $a$ is small~\cite{seige}: {\it i.e.} there are large cancellations
in Eq. \ref{eq:mzsHS}. 
By equating $m_Z^2$ in terms of weak scale parameters (Eq. \ref{eq:mzs}) with 
$m_Z^2$ in terms of high scale parameters Eq. \ref{eq:mzssugra}), 
and using the fact that $\mu (\Lambda )\simeq \mu (weak)$, then also 
$am_{3/2}^2\simeq m_{H_u}^2(weak)$ and so a low value of $\Delta_{BG}$ also requires a low value
of $m_{H_u}^2(weak) \sim -m_Z^2$. By properly evaluating BG fine-tuning in terms of independent SUGRA parameters, 
namely $m_{3/2}$ and $\mu (\Lambda )$, then we are lead back to the same sort of conditions as implied by
low $\Delta_{EW}$: {\it i.e.} that $m_{H_u}^2$ is driven radiatively to small negative values.
In this manner
\bi 
\item $\Delta_{EW}$ encompasses {\it both} high scale and weak scale fine-tuning.
\ei 
The ambiguity between fine-tuning measures is removed.\footnote{
Another common fine-tuning measure\cite{fathiggs,kitnom,oldnsusy} is known as Higgs mass or large log fine-tuning $\Delta_{HS}$.
In its usual implementation, $\Delta_{HS}$ requires the radiative correction 
$\delta m_{H_u}^2$ to the Higgs mass $m_h^2\simeq \mu^2+m_{H_u}^2(\Lambda )+\delta m_{H_u}^2$
be comparable to $m_h^2$. 
This contribution is usually written as $\delta m_{H_u}^2|_{rad}\sim -\frac{3f_t^2}{8\pi^2}(m_{Q_3}^2+m_{U_3}^2+A_t^2)\ln\left(\Lambda^2/m_{SUSY}^2 \right) $ which is used to claim that third generation squarks
$m_{\tst_{1,2},\tb_1}$ be approximately less than 500 GeV and $A_t$ be small for naturalness. 
This expression for $\delta m_{H_u}^2$ makes several approximations-- the worst of which is to neglect that
the value of $m_{H_u}^2$ itself contributes to $\delta m_{H_u}^2$. By combining dependent
contributions, then instead one requires that the two contributions 
$m_h^2 = \mu^2+\left( m_{H_u}^2(\Lambda) +\delta m_{H_u}^2\right)$
be comparable to $m_h^2$. Since $ m_{H_u}^2(\Lambda) +\delta m_{H_u}^2=m_{H_u}^2(weak)$, 
then we are lead back to the same 
conditions as low $\Delta_{EW}$ and low $\Delta_{BG}$.}
For electroweak naturalness in SUSY theories, it is not the case that sparticles
need to be near the scale $\sim 100$ GeV: it is just the weak scale 
Lagrangian parameters $m_{H_u}$ and $\mu$: the remaining soft breaking terms
may lie comfortably in the multi-TeV range at little cost to naturalness.

\section{How much fine-tuning is too much?}
\label{sec:30}

Once a reliable measure of fine-tuning is established, then the next question is: for which values of
$\Delta_{EW}$ is a model natural, and for which can it be considered fine-tuned? The original BG paper
considered $\Delta_{BG}<10$ to be natural. However, as experimental limits on sparticle masses grew,
then much higher values of $\Delta_{BG}$ were tolerated-- up to $\sim 100$\cite{feng} or even 
$\sim 1000$\cite{g_ross}. 
This increased tolerance perhaps reflected a reluctance to easily give up on an 
amazingly beautiful and simple  paradigm even in the face of apparent large fine-tuning.

\begin{figure}[tbp]
\includegraphics[height=0.4\textheight]{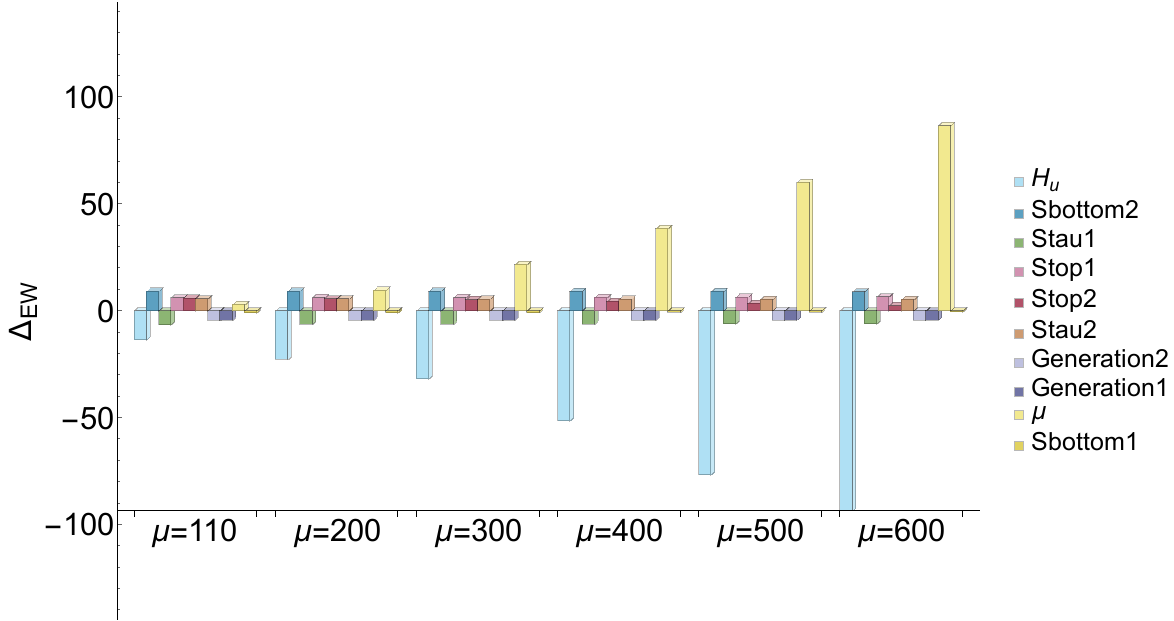}
\caption{Plot of contributions to $\Delta_{EW}$ for various values of
superpotential $\mu$ parameter.
\label{fig:comp}}
\end{figure}

By properly evaluating $\Delta_{BG}$ in terms of Eq. \ref{eq:mzssugra}, or equivalently $\Delta_{EW}$ 
(which includes one-loop radiative corrections), then we can re-evaluate how much fine-tuning is too much.
In Fig. \ref{fig:comp}, we show the dominant contributions to $\Delta_{EW}$ for a simple
NUHM2 benchmark model with $m_0=5$ TeV, $m_{1/2}=700$ GeV, $A_0=-8.3$ TeV and $\tan\beta =10$ with 
$m_A=1$ TeV and variable $\mu$. In the first case, with $\mu =110$ GeV and $\Delta_{EW}=13.6$, 
we display the several largest contributions to $\Delta_{EW}$. 
Some are positive and some are negative but all are comparable to $\pm 10$. 
This case is visually highly natural: the $Z$ mass is $\sim 100$ GeV because its various
contributions are $\sim 100$ GeV. In the second column, with $\mu =200$ GeV and $\Delta_{EW}=22.8$, 
the $H_u$ contribution to $\Delta_{EW}$ is largest but is on the whole balanced by several comparable 
positive terms: again, one would not claim it as unnatural. The third column with $\mu =300$ GeV, we see 
that the $H_u$ contribution to $\Delta_{EW}$ is again the largest, 
but in this case now a value of $\mu $ must be selected
as large positive to compensate and enforce that $m_Z=91.2$ GeV: 
this case is starting to already become unnatural.
By the fourth column with $\mu =400$ GeV and $\Delta_{EW}=51.5$, the fine-tuning is visually striking: 
the model is no longer natural. 
The unnaturalness is only accentuated in columns five and six for $\mu =500$ and 600 GeV respectively.

To be conservative, it is evident that fine-tuning has set in for $\mu$ values $\sim 300-400$ GeV.
For such cases one would expect the value of $m_Z$ also to be in the $300-400$ GeV range. 
With a value of $\mu=350$ GeV as a conservative upper estimate, 
and with the contribution $\Delta_{EW}(\mu )=\mu^2/(m_Z^2/2)$, 
we would then expect already values of $\Delta_{EW}>30$ to be overly fine-tuned. 
This is somewhat above the values expected in the original BG paper where they adopted $\Delta_{BG}^{max}=10$. 
If we increase the mass bounds from
BG by a factor $\sqrt{\Delta_{EW}^{max}/10}$ then we expect from BG with $m_t=173.2$ GeV 
that $\mu <350$ GeV and $m_{\tg}\alt 350$ GeV.

From scans over the popular mSUGRA/CMSSM\cite{msugra,cmssm} model, it is found that with 
$m_h=125\pm 2$ GeV, the lowest value of $\Delta_{EW}$ which can be obtained is $\sim 100$\cite{sug_ft,comp,seige}. 
In this case, the mSUGRA/CMSSM model is highly fine-tuned and is already ruled out.
In addition, an assortment of other models-- including mGMSB, mAMSB and mirage unification models--
are also ruled out\cite{seige}. 
What remains is a region of NUHM2 parameter space where small $\mu\sim 100-300$ GeV is 
allowed and where highly-mixed top squarks may live in the few TeV range. This region of 
NUHM2 parameter space is labelled as RNS, standing for models with radiatively-driven natural 
supersymmetry\cite{ltr,rns}. 
RNS models are characterized by light higgsinos with mass $\sim \mu\sim 100-300$ GeV. 
The LSP is a higgsino-like WIMP with a thermally-produced underabundance of dark matter.
However, solving the QCD sector naturalness problem via the axion\cite{peccei} 
leads also to axion dark matter\cite{sikivie}
so that one expects two dark matter particles: the axion along with a higgsino-like WIMP\cite{mixDM}.
Variants of this model with greater parameter freedom, such as non-universal
gaugino masses leading to natural SUSY with bino-like or wino-like LSPs-- are also allowed\cite{binowino}.

\section{Mass bounds from naturalness in NUHM2}
\label{sec:mass}

To derive sparticle mass bounds from the naturalness principle,
we will generate SUSY spectra using Isajet~\cite{isajet,isasugra} in the 
2-parameter non-universal Higgs model~\cite{nuhm2} (NUHM2) which allows for very 
low values of $\Delta_{EW}<10$.
The parameter space is given by
\be
m_0,\ m_{1/2},\ A_0,\ \tan\beta,\ \mu,\ m_A,\  \qquad {\rm (NUHM2)}. 
\label{eq:nuhm2}
\ee
The NUHM2 spectra and parameter spread versus $\Delta_{EW}$ were evaluated in 
Ref.~\cite{rns} but with $m_A$ restricted to $<1.5$ TeV and $m_{1/2}<2$ TeV. 
Here, we update these results by ensuring that we use sufficiently large range of input parameters that
our upper bounds on sparticle masses surely come from $\Delta_{EW}<30$ rather than from
artificial upper limits on scan parameters. 
Some previous mass bounds were extracted in Ref's \cite{rns,AbdusSalam:2013qba,AbdusSalam:2015uba}.
Here, we enlarge the scan region to include:
\bea
m_0 &:& \ 0-20\ {\rm TeV}, \nonumber\\
m_{1/2} &:& \  0.3-3\ {\rm TeV},\nonumber\\
-3 &<& A_0/m_0 \ <3,\nonumber\\
\mu &:& \ 0.1-1.5\ {\rm TeV}, \label{eq:param}\\
m_A &:& \ 0.15-20\ {\rm TeV},\nonumber\\
\tan\beta &:& 3-60 . \nonumber
\eea
We require of our solutions that:
\bi
 \item electroweak symmetry be radiatively broken (REWSB),
 \item the neutralino $\tz_1$ is the lightest MSSM particle,
 \item the light chargino mass obeys the model
independent LEP2 limit, $m_{\tw_1}>103.5$~GeV~\cite{lep2ino},
\item LHC8 search bounds on $m_{\tg}$ and $m_{\tq}$ 
from the $m_0$ vs. $m_{1/2}$ plane\cite{atlas_s} are respected,
\item $m_h=125\pm 2$~GeV.
\ei
It is important to note in this Section that some of our upper bounds come from the specific model we sample.
For instance, the extracted upper bound on the gluino mass comes from RG running effects where the gluino
mass feeds into the top squark soft terms, and the top-squark soft terms are mainly constrained
by the $\Sigma_u^u(\tst_{1,2})$ contributions to $\Delta_{EW}$. 
In this respect also, $\Delta_{EW}$ is sensitive to high scale effects since it measures if  a 
particular model defined by high scale parameters can generate the weak scale
characterized by $m_{W,Z,h}\sim 100$ GeV.
In the next Section, we will discuss
model-independent upper bounds which do not depend on high scale physics.

\begin{figure}[tbp]
\includegraphics[width=8cm,clip]{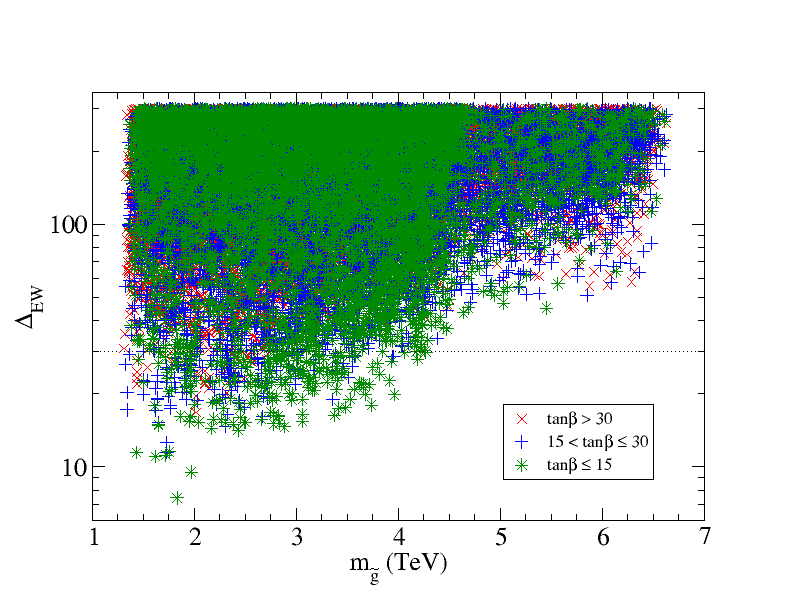}
\includegraphics[width=8cm,clip]{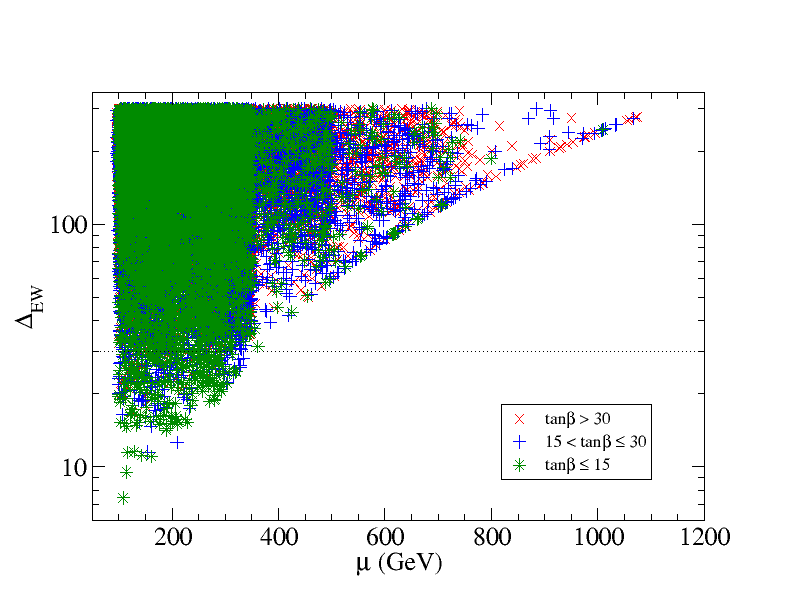}\\
\includegraphics[width=8cm,clip]{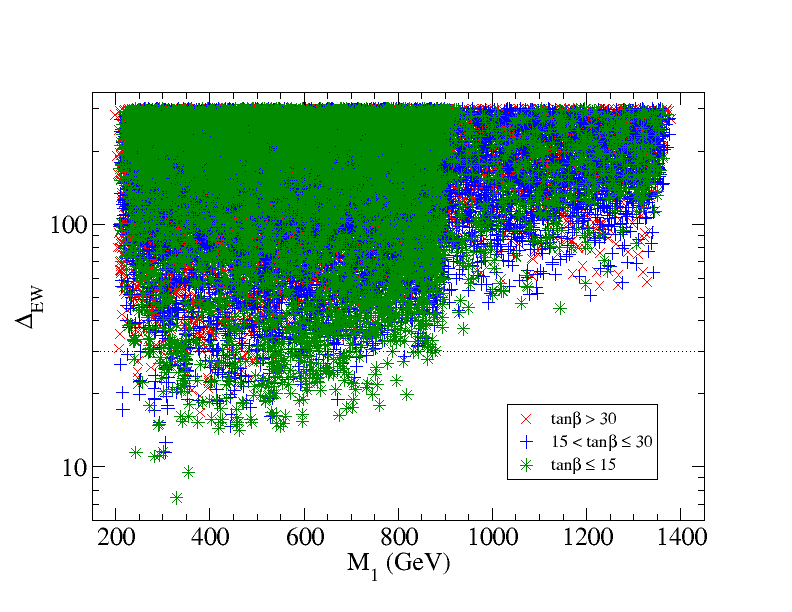}
\includegraphics[width=8cm,clip]{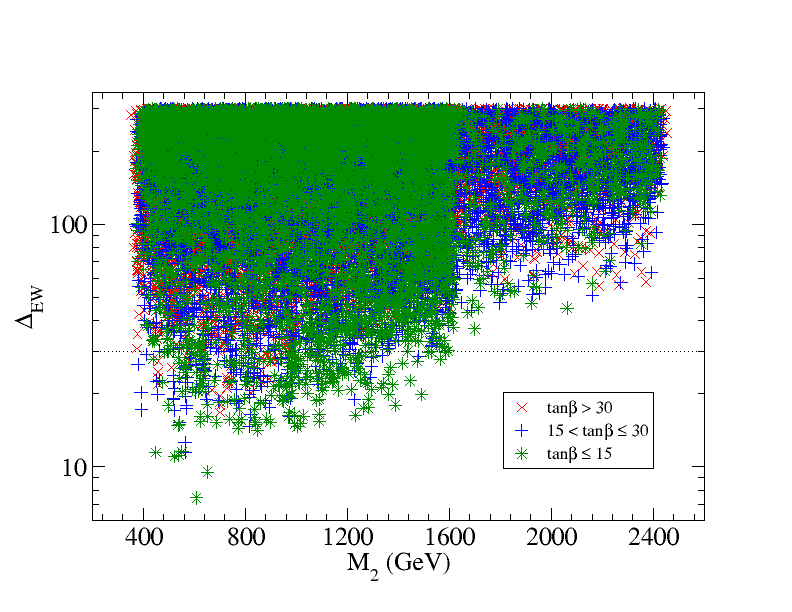}
\caption{Plot of $\Delta_{EW}$ vs. $m_{\tg}$, $\mu$, $M_1$ and $M_2$ 
from a scan over the NUHM2 model. Points with $\Delta_{EW}<30$ (below dotted line)
are considered natural.
\label{fig:m_inos}}
\end{figure}

The first results of our scan are shown in Fig. \ref{fig:m_inos}.
In frame {\it a}), we plot $\Delta_{EW}$ vs. $m_{\tg}$.
The symbols are color-coded according to low $(\le 15$), intermediate ($15-30$) and high ($>30$) values of $\tan\beta$.
From the plot, we see first that there is indeed an upper bound to $m_{\tg}$ provided by naturalness.
If we enforce $\Delta_{EW}<30$, then we find that 
\bi
\item $m_{\tg}\alt 4$ TeV. 
\ei
This is to be compared to the
reach of LHC for gluinos via gluino pair production followed by cascade decays. 
For the case of heavy squarks $m_{\tq}\gg m_{\tg}$ and with $\sim 1000$ fb$^{-1}$, 
LHC13 has a reach up to $m_{\tg}\sim 2$ TeV\cite{andre,lhc}. 
Thus, while natural SUSY may well be discovered by LHC13\footnote{For
an overview of what natural SUSY looks like at LHC13, see Ref.~\cite{baris}}, 
there is also plenty of natural SUSY parameter space well beyond the LHC reach for $\tg\tg$ production.
This bound may also be compared to the original BG result for $m_t=173.2$ GeV and with $\Delta_{BG}<30$:
there it was found that $m_{\tg}\alt 350$ GeV. The discrepancy between results arises because BG
evaluated $\Delta_{BG}$ within the four-parameter mSUGRA/CMSSM effective theory where 
$m_0$, $m_{1/2}$, $A_0$ and $B$ are assumed as independent. 
By recognizing the soft terms as dependent multiples of $m_{3/2}$, then
using Eq. \ref{eq:mzssugra} we have $\Delta_{BG}\simeq \Delta_{EW}$ 
and much larger values of $m_{\tg}$ are allowed while preserving naturalness. 
Our result may also be compared
with Feng\cite{feng} who uses an upper bound of 1\% fine-tuning in a multi-parameter SUSY 
effective theory: he then finds $m_{\tg}\alt 1.4$ TeV, 
slightly beyond the latest bound of $m_{\tg}\agt 1.3$ TeV from LHC8 searches.
If we scale this back to 3.3\% fine-tuning to compare with our result, then Feng would obtain
$m_{\tg}\alt 770$ GeV, well below the LHC8 lower limit on $m_{\tg}$.

In Fig. \ref{fig:m_inos}, we also show $\Delta_{EW}$ versus {\it b}) the higgsino mass $\mu$ and 
{\it c,d}) versus the gaugino masses $M_1$ and $M_2$. 
Since $m_{\tw_1}\sim m_{\tz_{1,2}}\sim |\mu |$ in models with a higgsino-like LSP, 
then we expect that 
\bi
\item $m_{\tw_1},\ m_{\tz_{1,2}}\alt 350$ GeV 
\ei
(the lighter the better).
While the higgsino-like electroweak-ino masses are necessarily not too far from $m_{Z,h}$, they are
notoriously difficult to see at LHC due to their compressed spectrum\cite{inoworld,lhc}. 
A possible way forward may be via $pp\to \tz_1\tz_2 g$ production followed by
$\tz_2\to\mu^+\mu^-\tz_1$. In this case, the hard gluon ISR serves as a trigger so that events containing 
soft dimuons may be visible above SM backgrounds\cite{kribs,bmt,chan}.
In contrast, the required light higgsinos
should be easily visible at an $e^+e^-$ collider operating with $\sqrt{s}>2\mu$\cite{inoworld,ilc} 
where also their masses and mixings can be extracted to high precision. 
In this, we are in accord with the conclusion of BG\cite{bg} that lepton colliders provide a more powerful probe 
of SUSY electroweak naturalness than do hadron colliders.

The bound extracted from frame {\it c}) is that the bino mass $M_1\alt 900$ GeV for $\Delta_{EW}<30$.
Since the NUHM2 model assumes gaugino mass unification, this translates to the
mass on the third lightest neutralino $\tz_3$ which is then mainly bino-like.
 
The bound on the wino mass $M_2$ shown in frame {\it d}) translates to a bound on 
the wino-like electroweakinos of $m_{\tw_2,\tz_4}\alt 1600$ GeV. 
As noted in Ref's \cite{lhcltr,lhc}, wino pair production
at LHC13 (via $\tw_2^\pm\tz_4$ and $\tw_2^+\tw_2^-$ production) forms the dominant {\it visible} reaction. 
In fact, the reach of LHC13 via the same-sign diboson (SSdB) signature 
($pp\to\tw_2^\pm\tz_4\to W^\pm W^\pm +\eslt$) exceeds the reach via gluino
pair production for integrated luminosities $\agt 300$ fb$^{-1}$. The reach of LHC13 for wino pairs via 
the SSdB signature extends to $m_{\tw_2}\alt 680$ GeV.

\begin{figure}[tbp]
\includegraphics[height=0.3\textheight]{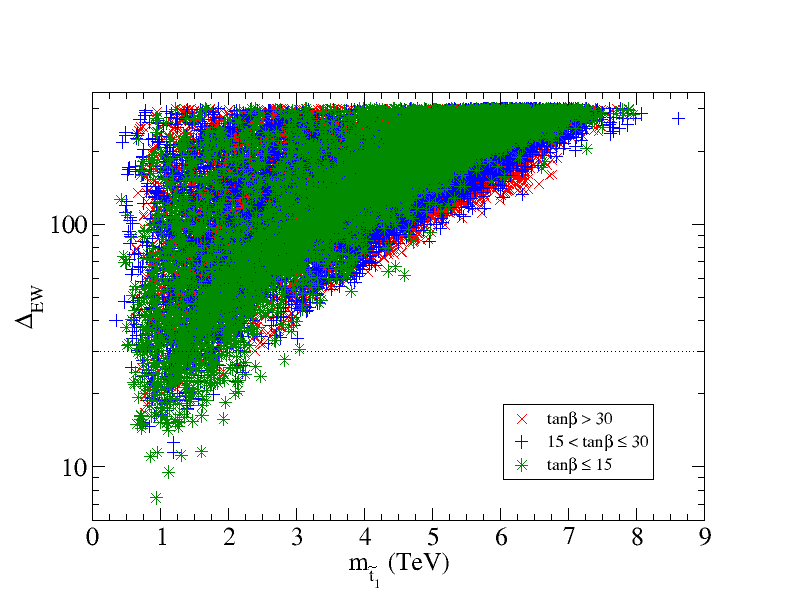}
\includegraphics[height=0.3\textheight]{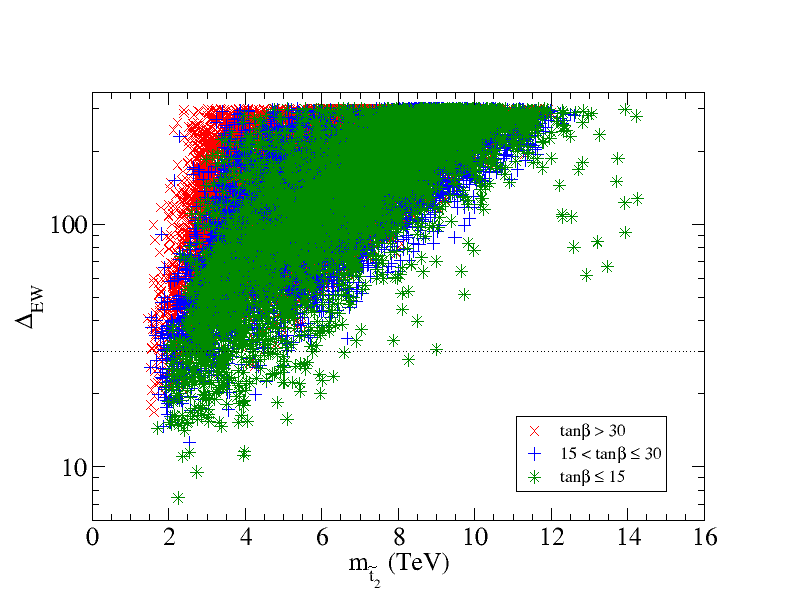}\\
\includegraphics[height=0.3\textheight]{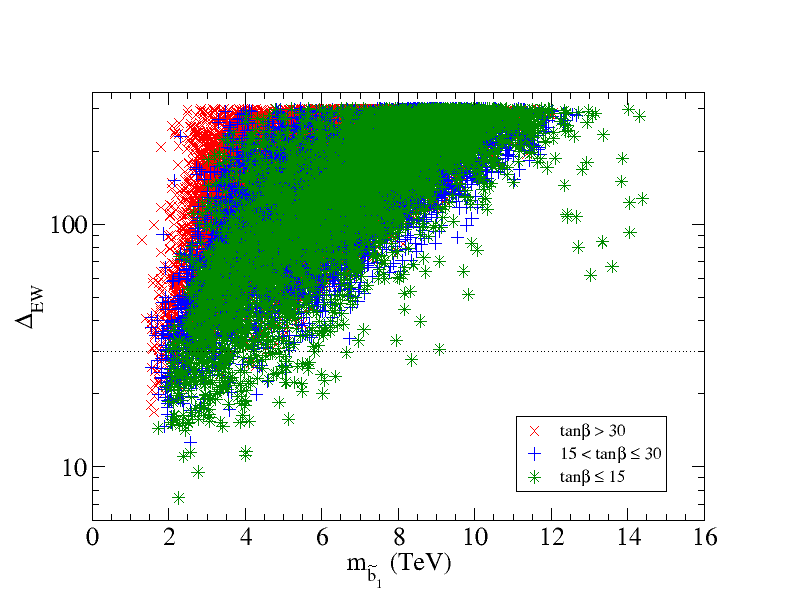}
\includegraphics[height=0.3\textheight]{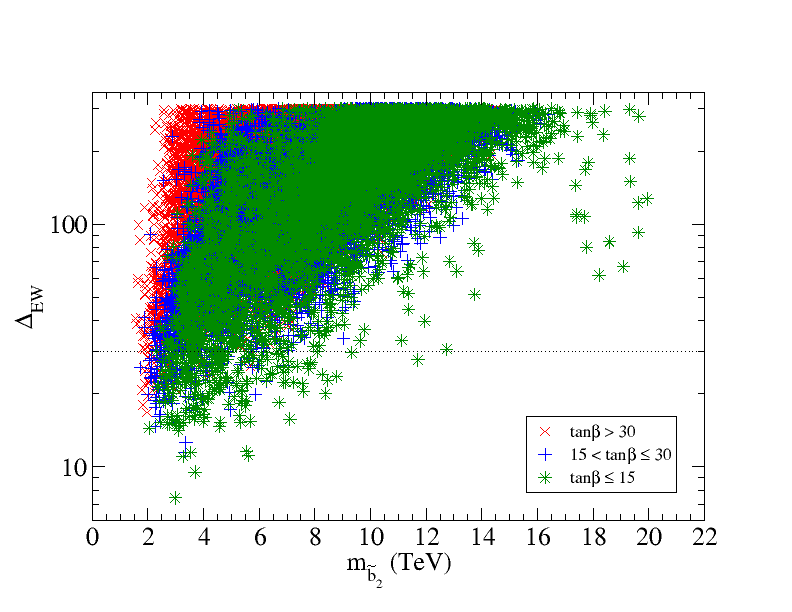}
\caption{Plot of $\Delta_{EW}$ vs. $m_{\tst_1}$, $m_{\tst_2}$, $m_{\tb_1}$ and $m_{\tb_2}$ 
from a scan over the NUHM2 model.
\label{fig:m_3rd}}
\end{figure}

In Fig. \ref{fig:m_3rd}, we show updated upper bounds on third generation squark masses.
From frame {\it a}), we see that 
\bi
\item $m_{\tst_1}\alt 3$ TeV. 
\ei
This upper bound is much higher than previous 
incarnations of natural SUSY based on large-log fine-tuning\cite{oldnsusy} (where it was claimed that
naturalness required three third generation squarks  with mass $m_{\tst_{1,2},\tb_1}\alt 600$ GeV). 
The projected reach of LHC13 for $m_{\tst_1}$ in various
simplified models extends up to $m_{\tst_1}\sim 1$ TeV. Thus, the lighter top squark may also lie well
beyond LHC13 search capabilities with little cost to naturalness.
From frames {\it b}), {\it c}) and {\it d}), we find successively that $m_{\tst_2}\alt 9$ TeV, $m_{\tb_1}\alt 9$ TeV and
$m_{\tb_2}\alt 10$ TeV.

\begin{figure}[tbp]
\includegraphics[width=8cm,clip]{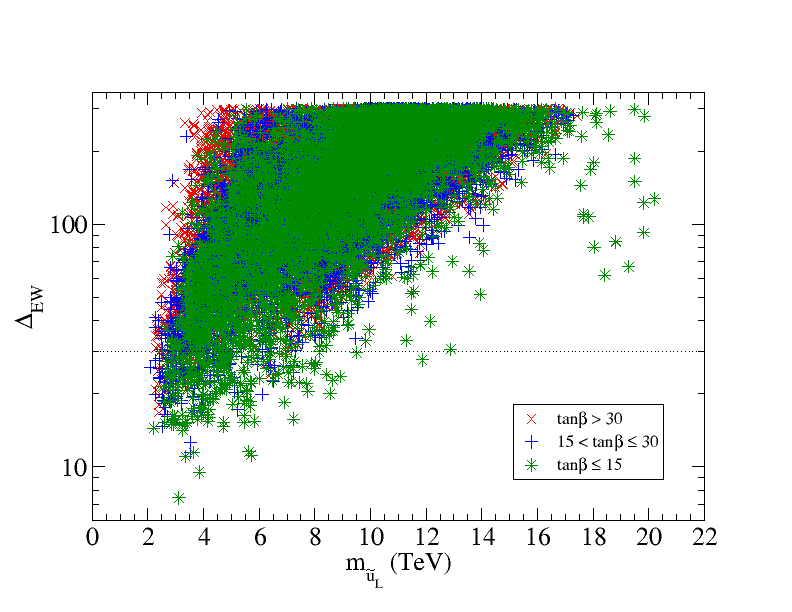}
\includegraphics[width=8cm,clip]{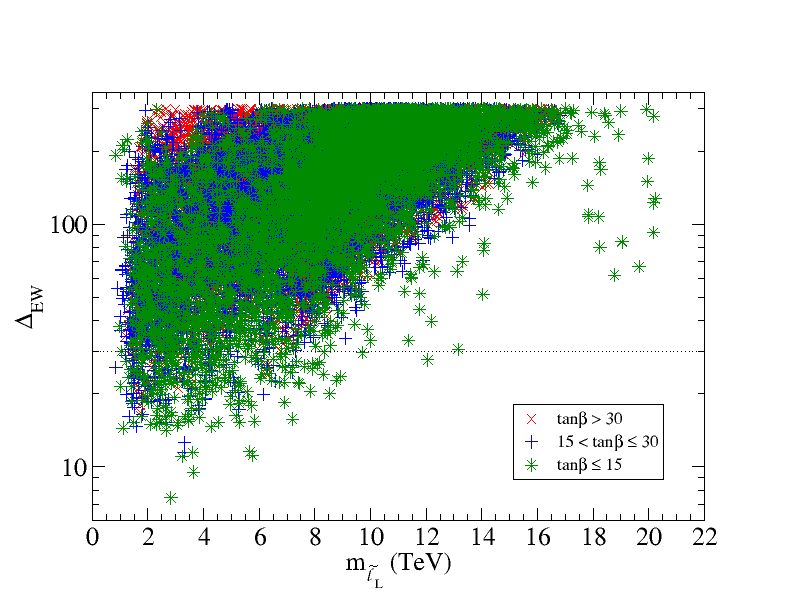}\\
\includegraphics[width=8cm,clip]{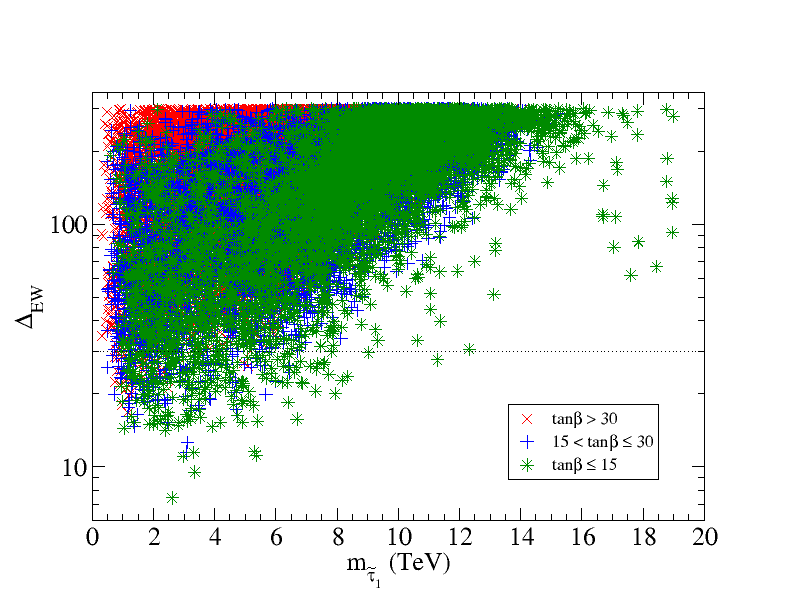}
\includegraphics[width=8cm,clip]{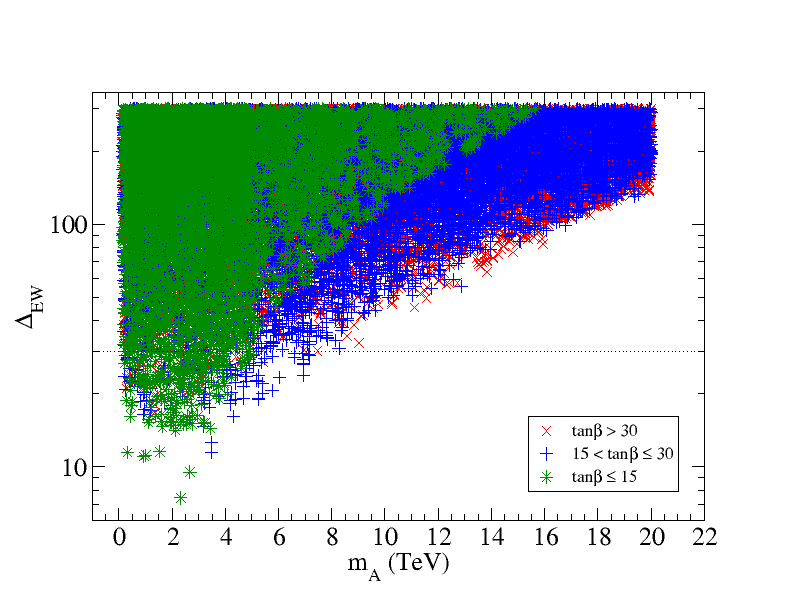}
\caption{Plot of $\Delta_{EW}$ vs. $m_{\tu_L}$, $m_{\tell_L}$, $m_{\ttau_1}$ 
and $m_{A}$ from a scan over the NUHM2 model.
\label{fig:m_scalars}}
\end{figure}

In Fig. \ref{fig:m_scalars} we show plots of $\Delta_{EW}$ versus various matter scalar masses
{\it a}) $m_{\tu_L}$, {\it b}) $m_{\tell_L}$, {\it c}) $m_{\ttau_1}$ and {\it d}) $m_A$.
Frame {\it a}) is typical of all first/second generation matter scalars. Here, we extract
an upper bound of $m_{\tu_L}\alt 10$ TeV. The bound on first/second generation matter scalars
arises from $D$-term contributions to $\Sigma_u^u$\cite{rns}. 
For certain mass degeneracy patterns listed in Ref. \cite{maren2}, 
these contributions nearly cancel amongst themselves. 
The limits become much stronger for non-degenerate matter scalars. The limits are also affected by
two-loop RG effects where heavy first/second generation matter scalars feed into third generation and Higgs 
soft term evolution. In this case, large $m_0(1,2)$ can drive $m_{H_u}^2$ to large
instead of small negative values\cite{Baer:2000xa}. The limit on first/second generation
sleptons, shown in frame {\it b}), is similar. If we allow for non-degenerate generations, {\it i.e.}
$m_0(1,2)\ne m_0(3)$, then the upper bounds on first/second generation squarks and sleptons
can increase to $\sim 20$ TeV\cite{rns}.

In frame {\it c}), we show $\Delta_{EW}$ vs. $m_{\ttau_1}$. In this case, the upper bound on third generation
sleptons is also $m_{\ttau_1}\alt 10$ TeV. 
An upper bound on $m_A$ can be extracted from frame {\it d}) of Fig. \ref{fig:m_scalars}. Here we find
$m_A\alt 5\ (8)$ TeV for $\tan\beta <15\ (50)$. Thus, the heavy Higgs bosons with mass
$m_H\sim m_{H^\pm}\sim m_A$ may also be well beyond the reach of LHC13 at little cost to naturalness.

\subsection{Comparison of original BG results to this work}

In Table \ref{tab:BGcomp}, we compare the upper limits on sparticle masses
and the $\mu$ parameter extracted from the original BG paper\cite{bg} 
as compared to this work.
The BG results are presented for $m_t=173.2$ GeV and scaled to impose
$\Delta_{BG} <30$ instead of $\Delta_{BG} <10$.
The BG results used $m_0$, $m_{1/2}$, $A_0$ and $\mu$ 
as independent parameters whereas our results combine these 
contributions to $m_Z^2$ since each is computed as a multiple of
$m_{3/2}$ in gravity mediation models. To include radiative corrections, 
we use the $\Delta_{EW}$ measure.

Since both groups use $\mu$ as an independent parameter, then both groups
agree on the upper bound $\mu\alt 350$ GeV leading to relatively light higgsinos: the closer to $m_h$ the better. 
However, for other sparticle masses, then our results differ markedly. 
Whereas BG find $m_{\tg}\alt 350$ GeV, we find $m_{\tg}\alt 4$ TeV-- possibly
well out of range of LHC13. 
Also, for the bino mass $M_1$, BG find $M_1\alt 90$ GeV while we find $M_1\alt 900$ GeV. 
For the wino mass $M_2$, BG find $M_2\alt 170$ GeV (leading to the suggestion that charginos
should appear at LEP2 if SUSY is not fine-tuned) 
while we find $M_2\alt 1700$ GeV (possibly out of range of LHC13 even via wino pair production 
with decays to same-sign dibosons\cite{lhcltr}).
Meanwhile, for matter scalars, BG found $m_{\tu_R}\alt 700$ GeV whereas we find
$m_{\tu_R}\alt 10$ TeV. If we allow for non-degenerate matter scalar generations
($m_0(1,2)\ne m_0(3)$) then our limit can increase to $\sim 20$ TeV\cite{rns}.
Likewise, BG found $m_{\te_R}\alt 520$ GeV whereas we expect $m_{\te_R}\alt 10$ TeV
(20 TeV for split families).
\begin{table}[!htb]
\renewcommand{\arraystretch}{1.2}
\begin{center}
\begin{tabular}{c|cc}
mass bound (GeV) & BG & this\ work  \\
\hline
$\mu$ & 350 & 350 \\
$m_{\tg}$ & 350 & 4000 \\
$M_1$ & 90  & 900 \\
$M_2$ & 170 & 1700 \\
$m_{u_R}$ & 700 & 10000\ (20000) \\
$m_{e_R}$ & 520 & 10000\ (20000) \\
\hline
\end{tabular}
\caption{Upper bounds on masses (in GeV) from naturalness with 
$\Delta <30$ from original BG paper compared to Sec. \ref{sec:mass} 
of this work in the NUHM2 model. The entries in parenthesis  would result if
one allows for non-degenerate generations of soft scalar masses 
$m_0(1,2)\ne m_0(3)$\cite{rns}.
}
\label{tab:BGcomp}
\end{center}
\end{table}

\section{Upper bounds from the 19 parameter pMSSM}
\label{sec:pmssm}

The mass bounds from the previous section depend on (reasonable) assumptions about high scale physics
which are implicit in the NUHM2 model. 
The previous NUHM2 bounds were derived from requiring that no contributions to the 
renormalization-group-improved one-loop scalar potential be far larger than $m_Z^2/2$.
The RG-improved scalar potential contains in fact leading two loop terms since 
the parameters entering the scalar potential contain the effects of RG running.
For instance, the bound on the gluino mass arises mainly from its RG contribution to the stop masses:
a large value of $M_3$ pushes $m_{\tst_{1,2}}$ to large values leading to large $\Sigma_u^u(\tst_{1,2})$
contributions to $m_Z$.
Likewise, other soft terms contribute to the evolution of $m_{H_u}^2$ to small negative values at the weak scale.

It is popular in recent years to dispense with RG running and examine physics within the
phenomenological MSSM, a model with 19 free weak scale parameters\cite{pmssm}. 
Results from the pMSSM will look in some cases very different from those 
obtained with high scale models even though
in some sense the pMSSM contains models like NUHM2, and even though a given spectra
generated within either of the NUHM2 or the pMSSM models will yield up exactly 
the same value of $\Delta_{EW}$.
The main difference is that the two-loop contribution to the scalar potential arising from
RG running of $M_3$ will no longer occur, thus obviating any bound on the gluino mass.
In this case, we add in the explicit leading two-loop terms of order $\alpha_t\alpha_s$
as computed by Dedes and Slavich\cite{slavich}. These terms contain sensitivity to 
$m_{\tg}$ and to $m_{\tst_{1,2}}$. 

To obtain mass bounds from the pMSSM, we implement a scan over the 
19-dimensional weak scale parameter space with the following limits:
\bi
\item $m_{\tst_L},\ m_{\tst_R},\ m_{\tb_R},\ m_{\ttau_L},\ m_{\ttau_R}:0.2-20$ TeV, 
\item $m_{\tu_L},\ m_{\td_R},\ m_{\td_R},\ m_{\te_L},\ m_{\te_R}:0.1-20$ TeV, 
\item $A_t,\ A_b,\ A_\tau:-40\to +40$ TeV, 
\item $M_1:0.05-10$ TeV, $M_2:0.1-10$ TeV, $M_3:0.4-10$ TeV, 
\item $\mu:100-500$ GeV, $m_A:0.15-20$ TeV,
\item $\tan\beta:3-60$. 
\ei
We actually input the pMSSM parameters at a scale $\Lambda_{pMSSM}=20$ TeV-- 
just above the maximal soft masses. This ensures the reduced scale dependence
of the scalar potential from the interplay between the 
RG running of the soft terms and the $\log(Q^2)$
dependence of the loop contributions $\Sigma_u^u(i)$ and $\Sigma_d^d(j)$.
The contributions of $\Sigma_d^d(i)$ to $\Delta_{EW}$ are
suppressed by $\tan^2\beta$ and so typically are of little consequence.

The pMSSM tree level bounds from $\Delta_{EW}<30$ can be read off directly from Eq. \ref{eq:mzs}
without the need for any scan:
\bi
\item $\mu \alt 350$ GeV,
\item $m_{H_u}(weak)\alt 350$ GeV,
\item $m_{H_d}\alt 350\ {\rm GeV} \tan\beta$.
\ei
For large $m_{H_d}$, then $m_{H_d}\simeq m_A$ so the latter bound translates to a bound on $m_A$.

\begin{figure}[tbp]
\includegraphics[width=8cm,clip]{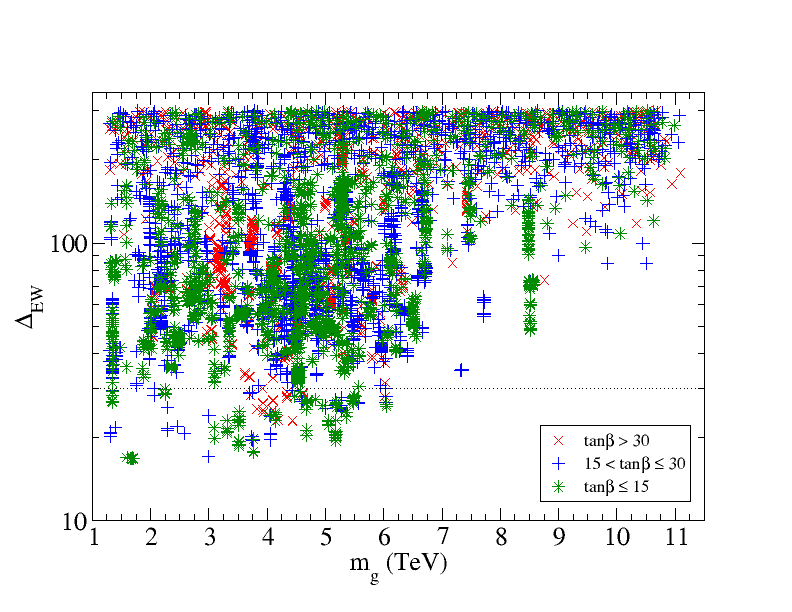}
\includegraphics[width=8cm,clip]{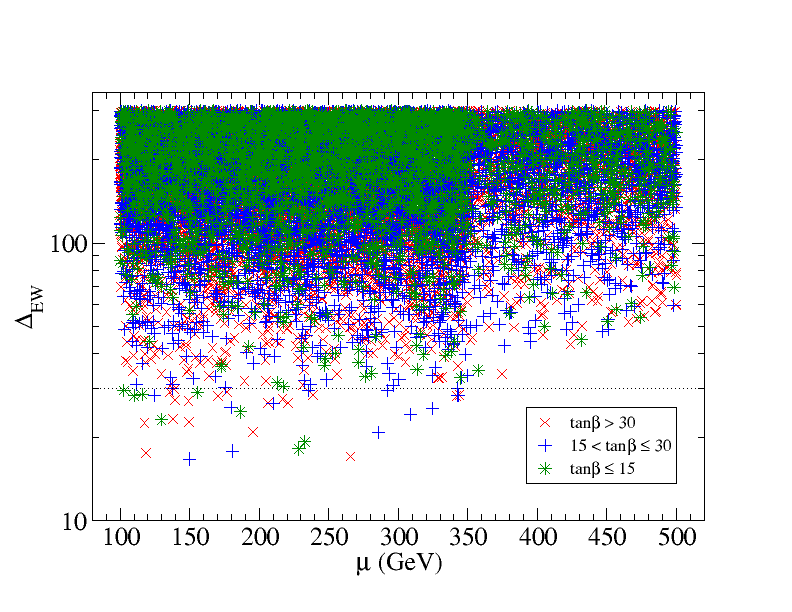}\\
\includegraphics[width=8cm,clip]{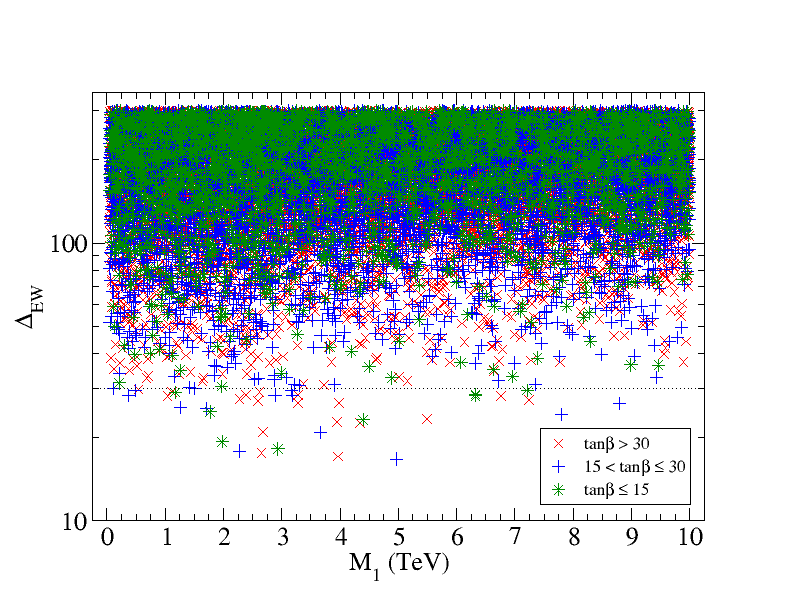}
\includegraphics[width=8cm,clip]{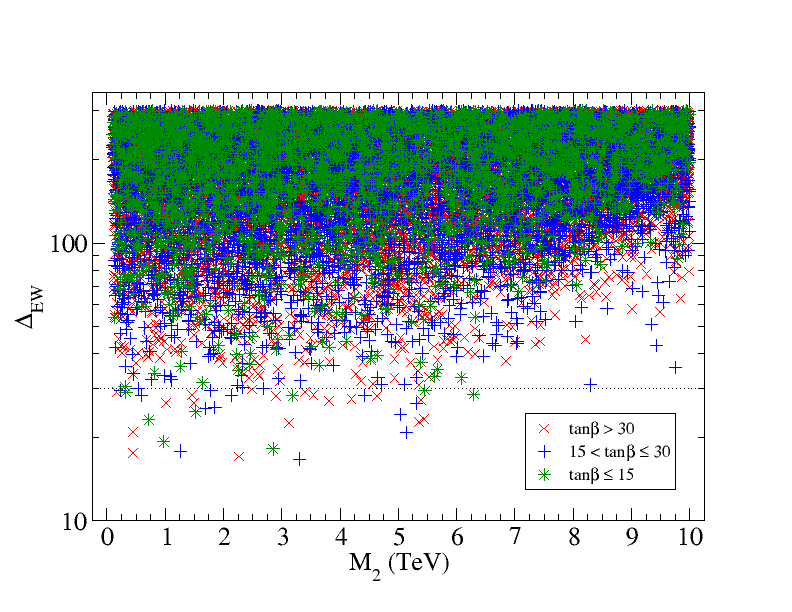}
\caption{Plot of $\Delta_{EW}$ vs. $m_{\tg}$, $\mu$, $M_1$ and $M_2$ 
from a scan over the 19 weak scale pMSSM parameters. 
Points with $\Delta_{EW}<30$ (below dotted line) are considered natural.
\label{fig:pmssm_inos}}
\end{figure}

In Fig. \ref{fig:pmssm_inos}, we plot the value of $\Delta_{EW}$ vs. the same parameters
as in Fig. \ref{fig:m_inos}: {\it a}) $m_{\tg}$, {\it b}) $\mu$, {\it c}) $M_1$ and {\it d}) $M_2$.
From frame {\it a}) we see a very important result: in the pMSSM, the bound on 
$m_{\tg}$ for $\Delta_{EW}<30$ has moved up to about 7 TeV, 
well beyond what was obtained for the NUHM2 model.
This bound arises after the inclusion of the order $\alpha_t\alpha_s$ two-loop 
contribution to the scalar potential which includes sensitivity to $m_{\tg}$\cite{slavich}.

From frame {\it b}), we see that, as expected, $\mu$ is once again bounded by $\mu\alt 350$ GeV 
for $\Delta_{EW}<30$, as is required by the tree-level contribution to Eq. \ref{eq:mzs}.
In frame {\it c}), we plot $\Delta_{EW}$ vs. the bino mass $M_1$. For the pMSSM, mass bounds on $M_1$
arise from the $\Sigma_u^u(\tz_i)$. Here, we see that $M_1$ can range as high as $9$ TeV-- far beyond the bounds
obtained in the NUHM2 model where gaugino mass unification is requred. In that case, the upper limit
on $m_{\tg}\alt 4$ TeV translates to a bound on $M_1\alt 0.8$ TeV. Likewise, in frame {\it d})
we plot $\Delta_{EW}$ vs. wino mass $M_2$. In this case, we find $M_2\alt 6$ TeV. This bound mainly arises from
the contributions $\Sigma_u^u(\tw_i)$. It is also far beyond the mass bound from NUHM2 where
$M_2\alt 1.6$ TeV was found.

\begin{figure}[tbp]
\includegraphics[height=0.3\textheight]{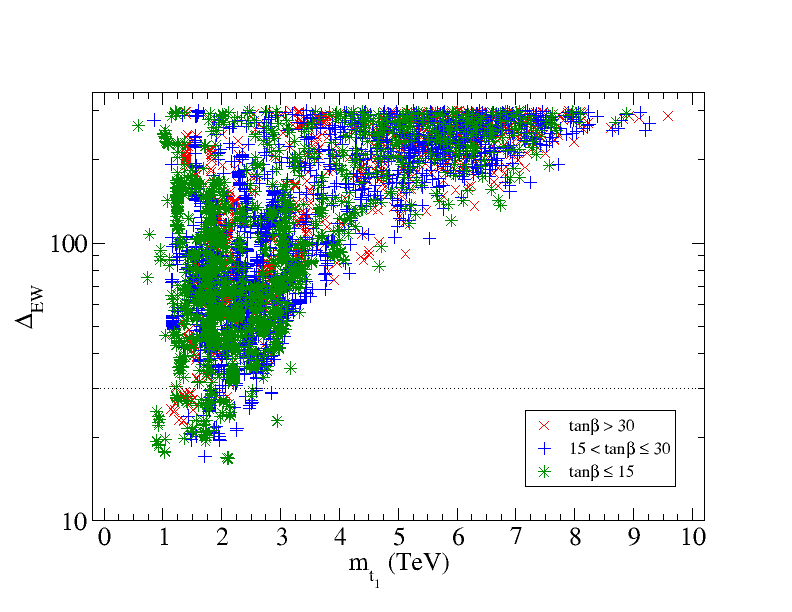}
\includegraphics[height=0.3\textheight]{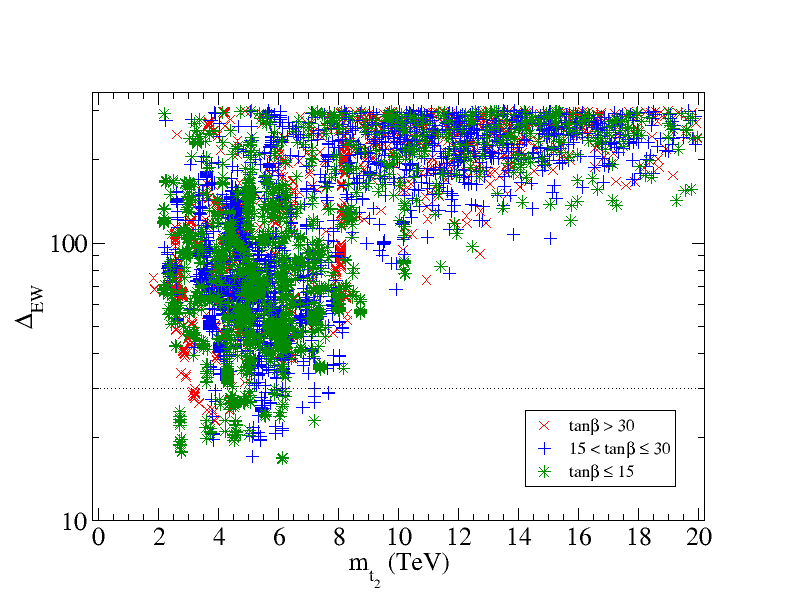}\\
\includegraphics[height=0.3\textheight]{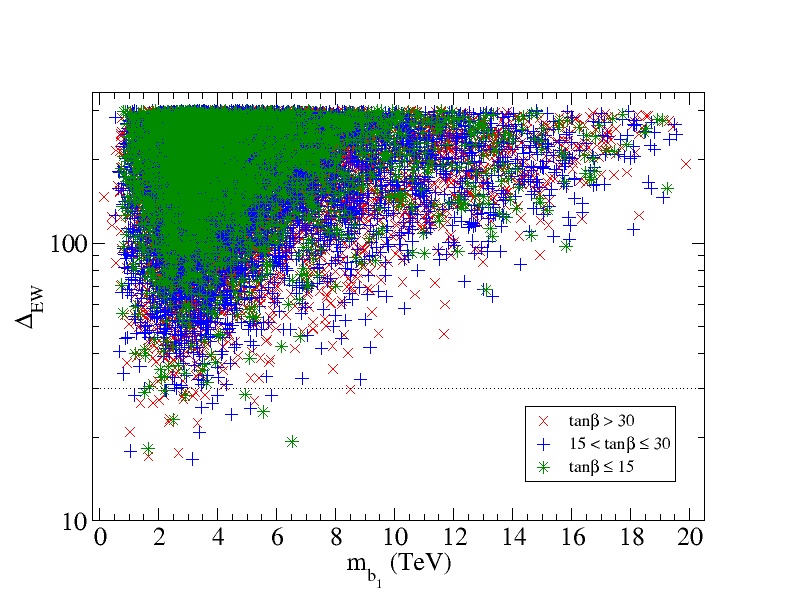}
\includegraphics[height=0.3\textheight]{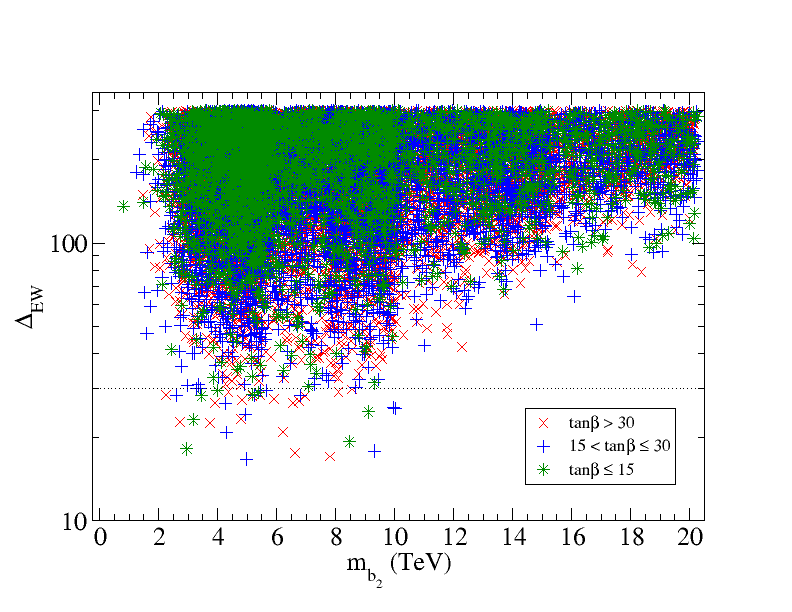}
\caption{Plot of $\Delta_{EW}$ vs. $m_{\tst_1}$, $m_{\tst_2}$, $m_{\tb_1}$ and $m_{\tb_2}$ 
from a scan over 19 weak scale pMSSM parameters.
\label{fig:pmssm_3rd}}
\end{figure}

In Fig. \ref{fig:pmssm_3rd} we display values of $\Delta_{EW}$ vs. 
{\it a}) $m_{\tst_1}$, {\it b}) $m_{\tst_2}$, {\it c}) $m_{\tb_1}$ and {\it d}) $m_{\tb_2}$.
From frame {\it a}), we find that $m_{\tst_1}\alt 3.5$ TeV, just slightly larger than the bound arising in the NUHM2 
model. This bound arises mainly due to the contributions $\Sigma_u^u(\tst_{1,2})$ to the scalar potential.
In frame {\it b}), we find that $m_{\tst_2}\alt 10$ TeV. This again is slightly beyond the bound
arising from the NUHM2 case with unified matter scalars at $m_{GUT}$.
From frames {\it c}) and {\it d}), we find that $m_{\tb_1}\alt 9$ TeV and $m_{\tb_2}\alt 10$ TeV. 
These results are in rough accord with values obtained from the NUHM2 model.

\begin{figure}[tbp]
\includegraphics[width=8cm,clip]{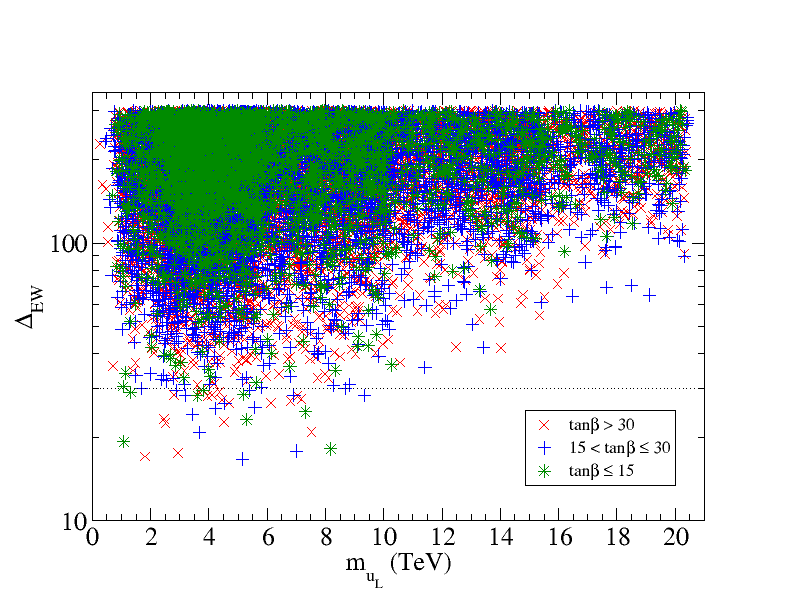}
\includegraphics[width=8cm,clip]{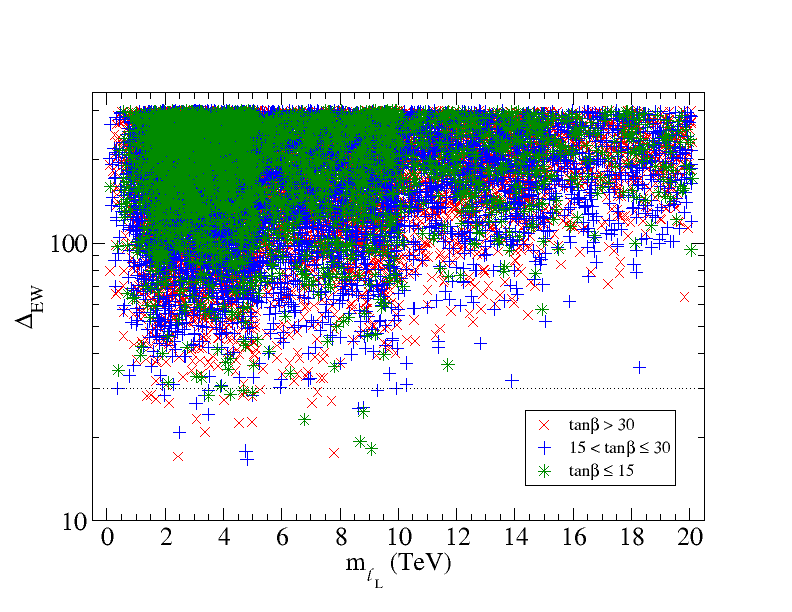}\\
\includegraphics[width=8cm,clip]{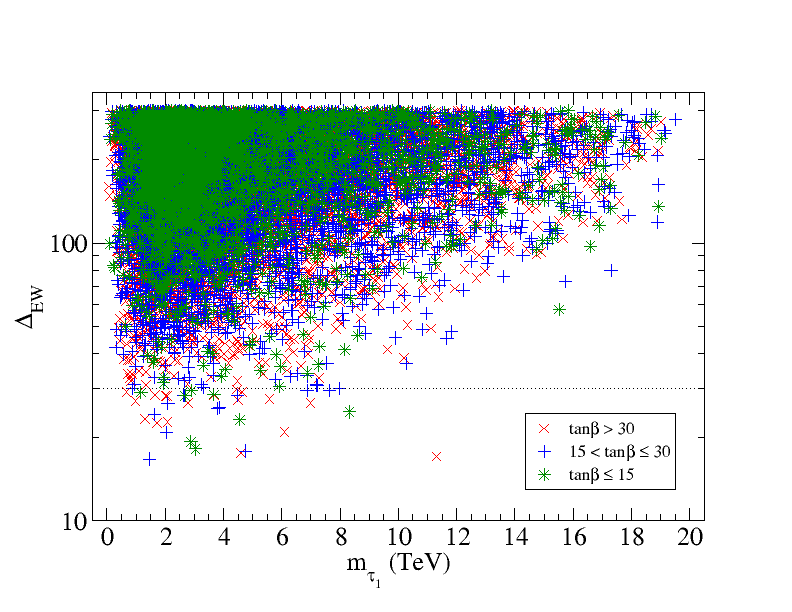}
\includegraphics[width=8cm,clip]{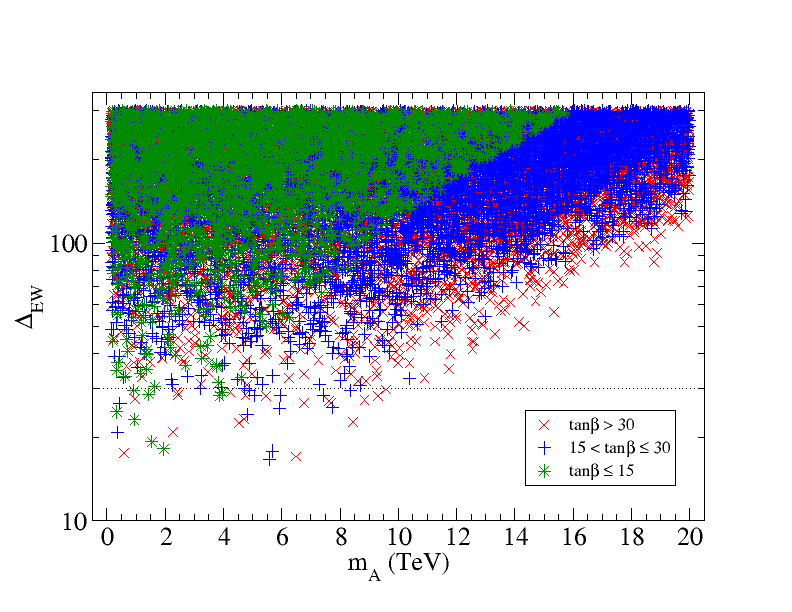}
\caption{Plot of $\Delta_{EW}$ vs. $m_{\tu_L}$, $m_{\tell_L}$, $m_{\ttau_1}$ 
and $m_{A}$ from a scan over 19 weak scale pMSSM parameters.
\label{fig:pmssm_scalars}}
\end{figure}

In Fig. \ref{fig:pmssm_scalars}, we attempt to extract upper mass bounds on squarks, sleptons, 
staus and heavy Higgs bosons in analogy to Fig. \ref{fig:m_scalars}. From frame {\it a}), we find naively that 
$m_{\tu_L}\alt 10$ TeV. This mass bound arises from $D$-term contributions to first/second generation 
scalar masses that enter the scalar potential via $\Sigma_u^u(\tu_L)$. 
As pointed out in Ref. \cite{maren2}, these $D$-term contributions  all cancel amongst themselves 
{\it provided that one of several mass degeneracy patterns exist}: 1. separately squark and slepton mass degeneracy, 
2. separately left- and right- sfermion degeneracy, 3. degeneracy within $SU(5)$ multiplets and
4. degeneracy within an entire generation, as expected in $SO(10)$ GUTs.
In these cases, the contributions-- which are all proportional to weak isospin and hypercharge assignments--
necessarily sum to zero for degenerate masses. When these contributions thus sum to zero, then
there are no bounds on first/second generation squark and slepton masses.
However, since it is highly improbable to generate these degeneracy patterns from a random pMSSM scan, then
mass bounds do arise from our random scan. As seen from frame {\it a}), we expect $m_{\tu_L}\alt 10$ TeV.
But this mass bound would disappear if we invoked degeneracy conditions amongst the physical masses within any of the patterns listed above. Likewise, in frame {\it b}), we see that $m_{\tell_L}\alt 10$ TeV and from frame {\it c}) we
expect $m_{\ttau_1}\alt 9$ TeV. The mass bound on $m_A$ which arises from the loop contributions $\Sigma_u^u(h,H,H^\pm )$
are shown in frame {\it d}). These constraints are actually somewhat stronger than the naive tree-level constraint
of $m_A\alt 350\ {\rm GeV}\tan\beta$ which is $\sim 18$ TeV for large $\tan\beta\sim 50$. Instead, we find
$m_A\alt 10$ TeV from our scan over 19 weak scale pMSSM parameters.

\subsection{Comparison between pMSSM and SUGRA19 model}

It may be worthwhile to compare the preceding results from the 19 parameter pMSSM 
model-- with inputs defined at the weak scale-- with results from Ref. \cite{sug19} where naturalness
was examined in the context of the SUGRA19 model with 19 input parameters defined at the GUT scale. 
The goal of Ref. \cite{sug19} was to see how how low in $\Delta_{EW}$ one might go 
within the context of a general SUGRA model, but not to establish upper bounds 
(which require a thorough rather than a focussed parameter space scan).
To compare sparticle mass upper bounds between pMSSM and SUGRA19, we have re-run SUGRA19 including
the aforementioned two-loop $\Sigma_u^u$ corrections with a thorough parameter space scan.
In both the pMSSM model and the SUGRA19 model, the SUSY mu parameter is bounded as $\mu< 350$ GeV for
$\Delta_{EW}<30$ since this quantity enters the scalar potential at tree-level.

\begin{figure}[tbp]
\includegraphics[width=8cm,clip]{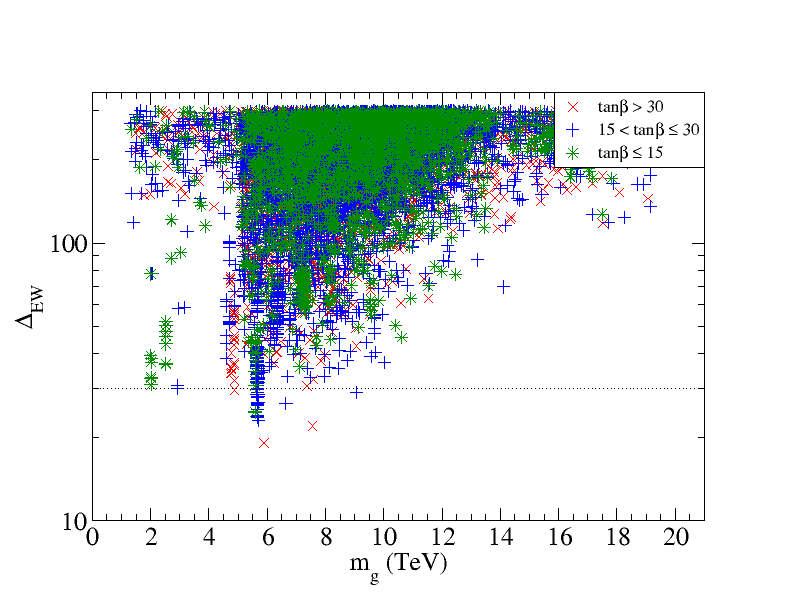}
\includegraphics[width=8cm,clip]{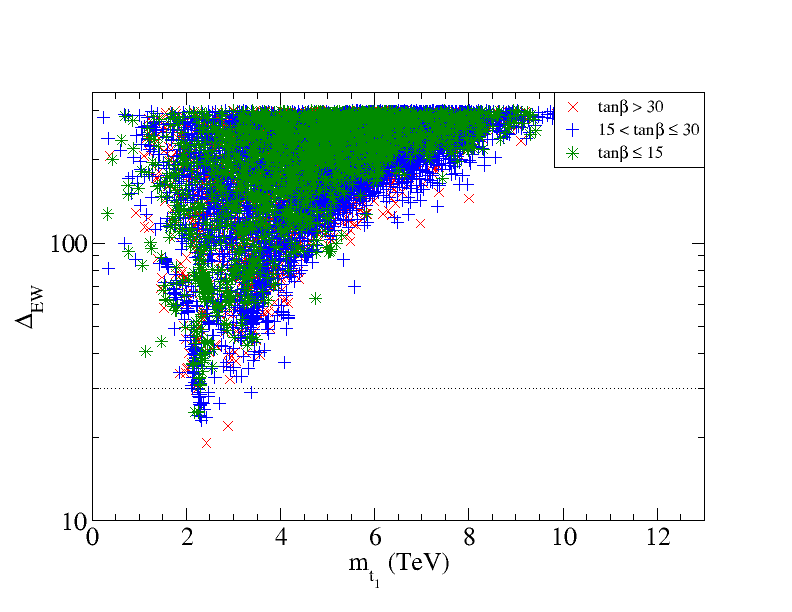}\\
\includegraphics[width=8cm,clip]{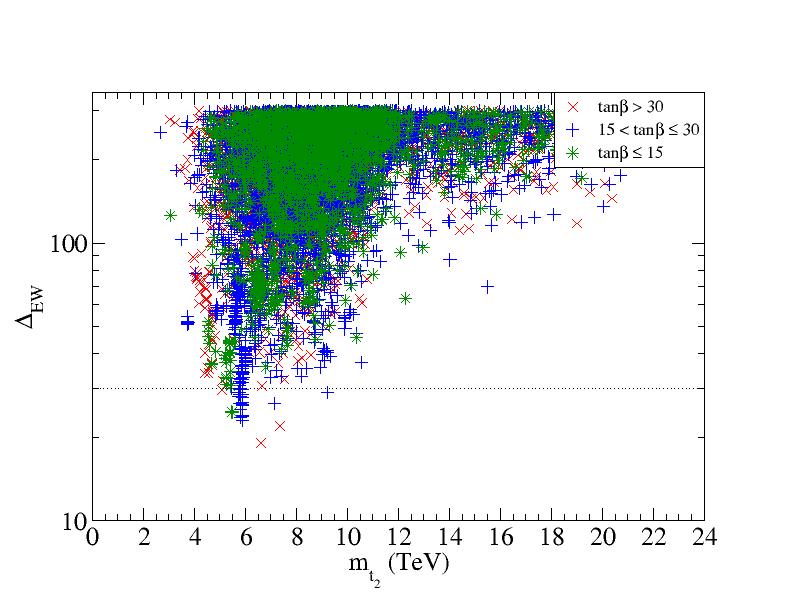}
\includegraphics[width=8cm,clip]{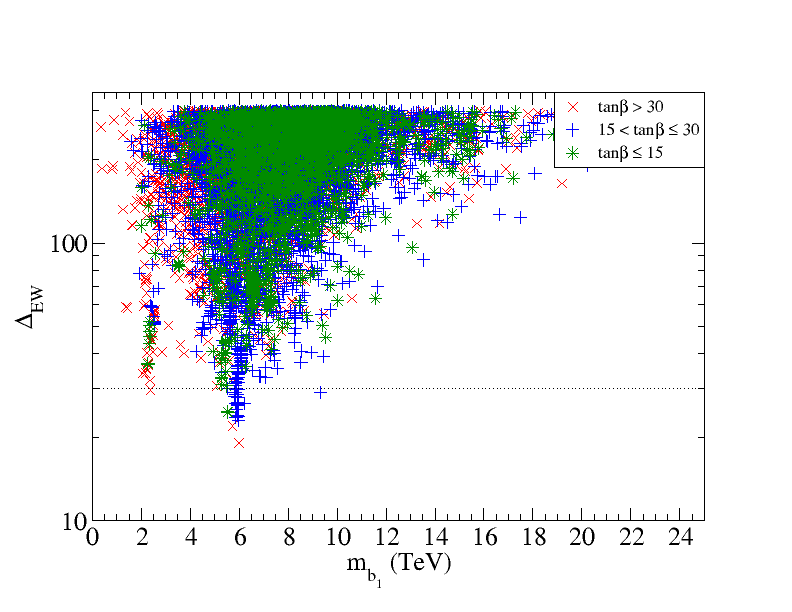}
\caption{Plot of $\Delta_{EW}$ vs. $m_{\tg}$, $m_{\tst_1}$, $m_{\tst_2}$ 
and $m_{\tb_1}$ from a scan over the SUGRA19 model.
\label{fig:sug19}}
\end{figure}

The results from the SUGRA19 model are shown in Fig. \ref{fig:sug19} for $\Delta_{EW}$ versus
{\it a}) $m_{\tg}$, {\it b}) $m_{\tst_1}$, {\it c}) $m_{\tst_2}$ and {\it d}) $m_{\tb_1}$.
The upper bound on the gluino mass is very similar in the two cases where we
find $m_{\tg}\alt 7$ TeV. 
Also, for both the SUGRA19 and pMSSM models, the third generation squark masses have similar upper bounds: 
$m_{\tst_1}\alt 3-3.5$ TeV and $m_{\tst_2,\tb_2}\alt 8-10$ TeV.  

From this comparison, and in the spirit of providing a target for experimenters seeking to 
verify or disprove SUSY, we would conclude that a negative search for light higgsinos with mass
$\alt 350$ GeV would rule out naturalness in the context of the MSSM 
(more complicated beyond-the-MSSM extensions could always
be built to circumvent these bounds\cite{nelson,luty,spmartin}). 
Thus, non-observation of light higgsinos would rule out both 
high scale and weak scale renditions of SUSY based on the MSSM. Alternatively, failure to find a
gluino with mass $m_{\tg}\alt 4$ TeV would rule out constrained SUSY models such as 
NUHM2 with unified soft parameters defined as high as $Q=m_{GUT}$. 
Colliders with the ability to probe beyond $m_{\tg}\sim 7$ TeV
would be needed to test/exclude gluino pair production within the pMSSM or SUGRA19 context\cite{pp100TeV}.

\section{Conclusions:} 
\label{sec:conclude}

In this paper, our goal was to sharpen the upper bounds on sparticle masses
arising from naturalness in order to provide a target for 
experimenters seeking to confirm or rule out the weak scale supersymmetry
hypothesis. 

While most sparticle search results are presented as lower bounds
in simplified or complete model parameter space, the question arises:
how far out in parameter space ought one to go before discovering weak scale
SUSY or claiming it is dead? While earlier papers by others have presented 
naturalness as a subjective, fuzzy and model-dependent notion,
instead here we argue that naturalness is 
\bi
\item objective, 
\item model-independent in that different models giving rise to 
the same spectra have the same value of naturalness, and 
\item predictive.
\ei
The previous confusion on this subject arose from what constitutes
independent model parameters. In the case of gravity mediation, 
for any given hidden sector the soft terms are calculable as multiples of the gravitino
mass $m_{3/2}$, {\it i.e.} they are {\it dependent}. By appropriately
combining dependent terms, then the BG measure implies the
same general consequences as the model {\it independent} electroweak measure $\Delta_{EW}$. 
We show visually that fine-tuning already arises at $\Delta_{EW}\sim 20-30$.
To be conservative, we take $\Delta_{EW}<30$ to derive upper bounds
on parameters and sparticle masses.

\begin{figure}[tbp]
\includegraphics[height=0.4\textheight]{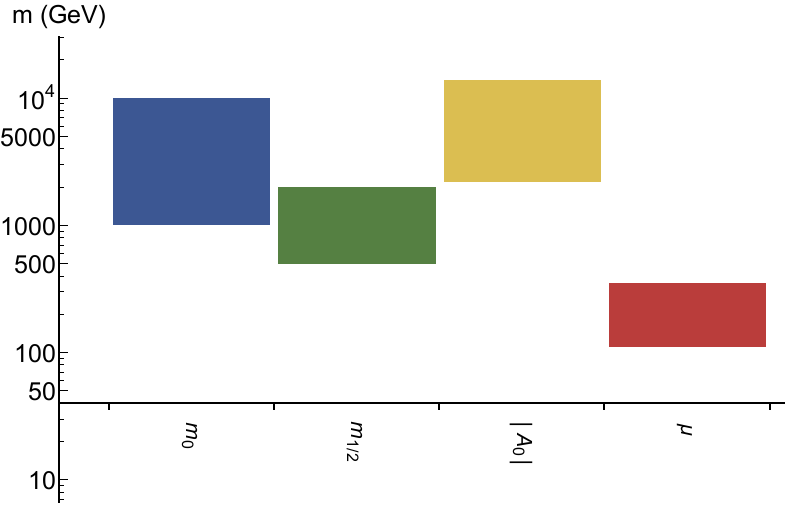}
\caption{Range of NUHM2 model parameters allowed by naturalness
with $\Delta_{EW}<30$.
\label{fig:bar_inputs}}
\end{figure}

In Sec. \ref{sec:mass}, we sharpened up previous bounds on sparticle masses
by increasing the range of parameters enough to ensure that upper bounds 
arose from $\Delta_{EW}<30$ and not from some artificial cut-off imposed
on our scan limits. For the NUHM2 model which allows SUGRA grand unification
and values of $\Delta_{EW}$ below 10, we found $m_{\tg}\alt 4$ TeV, well beyond
the reach of even high luminosity LHC. 
This bound is much higher than previous estimates which assumed that the various
soft SUSY breaking terms are independent.
Alternatively, we find the superpotential mu term to be $\mu\alt 350$ GeV. 
The range of NUHM2 input parameters which are allowed is shown in 
Fig. \ref{fig:bar_inputs}.

\begin{figure}[tbp]
\includegraphics[height=0.4\textheight]{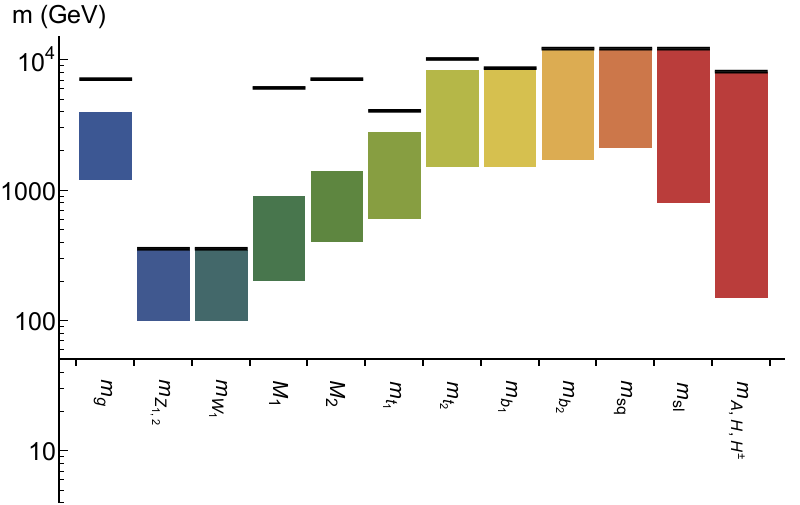}
\caption{Range of sparticle masses allowed by naturalness 
within NUHM2 model with $\Delta_{EW}<30$. The black bars show upper bounds from 
the pMSSM model with 19 weak scale parameters.
\label{fig:bar}}
\end{figure}

The allowed ranges of sparticle masses from NUHM2 are shown in Fig. \ref{fig:bar}
as the colored histograms.
As remarked before, the all-important gluino mass can range from
$m_{\tg}\sim 1.3-4$ TeV. Third generation squarks, which were argued
to exist in the sub-TeV regime in previous natural SUSY papers, are 
found to be consistent with naturalness when $m_{\tst_1}\alt 3$ TeV while
$m_{\tst_2,\tb_1}\alt 9$ TeV and $m_{\tb_2}\alt 10$ TeV. 
A large $A_t$ parameter here acts to reduce EW fine-tuning 
whilst lifting $m_h$ up to $\sim 125$ GeV.
The manifestation of large $A_t$ is a splitting of the top-squark mass eigenstates.
First and second generation sfermions can have masses ranging into the
10 TeV regime so long as they are sufficiently degenerate that 
naturalness contributions from $D$-terms largely cancel amongst themselves\cite{maren2}.
Such heavy matter scalars provide a decoupling solution to the SUSY
flavor and CP problems\cite{dine} and if their mass is comparable to $m_{3/2}$, then
one expects a decoupling solution to the gravitino problem as well.
The heavy Higgs bosons $A,\ H$ and $H^\pm$ can range up to the 5-8 TeV level.
While such heavy sparticle and Higgs masses are consistent with naturalness, 
typically we do not expect large deviations from SM rates 
in rare decay branching fraction measurements  such as
$B_s\to \mu^+\mu^-$ or $b\to s\gamma$ from natural SUSY\cite{rns} 
due to the presence of heavy mediators.

We have also extracted mass bounds on sparticles by requiring $\Delta_{EW}<30$ in 
a scan over the 19 dimensional weak scale parameter set of the pMSSM.
These bounds are shown as black lines in Fig. \ref{fig:bar} or as arrows when no bound arises.
The pMSSM bounds on scalar masses tend to be comparable to those from the NUHM2 model. However, bounds
on gaugino masses are severely different. By including leading two-loop contributions to the 
scalar potential, we find a bound of $m_{\tg}<7$ TeV arises in the pMSSM. 
Also, the bounds on the bino mass $M_1\alt 9$ TeV and 
wino mass $M_2\alt 6$ TeV-- which arise from neutralino and chargino loops-- are much higher than 
the corresponding upper bounds extracted from the NUHM2 model. 
A comparison of upper bounds extracted from NUHM2 and pMSSM is listed in Table \ref{tab:pmssm}.
\begin{table}[!htb]
\renewcommand{\arraystretch}{1.2}
\begin{center}
\begin{tabular}{c|cc}
mass & NUHM2 & pMSSM  \\
\hline
$\mu$    & 0.35 & 0.35 \\
$m_{H_u}(weak)$    & 0.35 & 0.35 \\
$m_A$ & 5-8 & 10 \\
$M_1$ & 0.9  & 9 \\
$M_2$ & 1.6 & 6 \\
$m_{\tg}$ & 4 & 7 \\
$m_{\tst_1}$ & 3 & 3 \\
$m_{\tst_2}$ & 9 & 9 \\
$m_{\tb_1}$ & 9 & 9 \\
$m_{\tb_2}$ & 10 & 10 \\
$m_{\tq}$ & 10 (20) & 10/none \\
$m_{\tell}$ & 10 (20) & 10/none \\
\hline
\end{tabular}
\caption{Upper bounds on masses (in TeV) from naturalness with 
$\Delta_{EW} <30$ from a scan over NUHM2 model versus a scan over 
the 19 weak scale parameter pMSSM.
The entries in parentheses  would result if
one allows for non-degenerate generations of soft scalar masses 
$m_0(1,2)\ne m_0(3)$\cite{rns}. The lack of bounds after the slash symbol
arise in the case where highly degenerate squark and slepton masses develop\cite{maren2}.
}
\label{tab:pmssm}
\end{center}
\end{table}

Our results have important implications for future particle physics facilities. 
Even the high luminosity LHC can explore only about half of natural SUSY parameter space. 
However, if we require a more stringent naturalness condition of $\Delta_{EW}\alt 10$
then $m_{\tg}\alt 2$ TeV and the {\it most natural} region of parameter space should be accessible to LHC13
(LHC13 has a projected $5\sigma$ discovery reach to $m_{\tg}\sim 2$ TeV 
for 300-1000 fb$^{-1}$ of data\cite{andre}).

The key feature of naturalness-- that quasi-degenerate higgsinos lie in the 100-350 GeV mass range-- 
highly motivates the construction of an $e^+e^-$ collider which can operate with $\sqrt{s}>2\mu$. 
Such a machine, constructed initially as a Higgs factory, would
turn out to be also a {\it higgsino factory} which would usher in the era of SUSY
discovery while simultaneously elucidating the nature of dark matter.
In this case, we  would expect it to consist of an admixture of higgsino-like WIMPs and
axions\cite{mixDM}.

\section*{Acknowledgments}

This work was supported in part by the US Department of Energy, Office of High Energy Physics.

\section{Appendix}
\label{sec:appendix}

In this appendix, we present some details about the radiative contributions $\Sigma_u^u(i)$
to the naturalness measure $\Delta_{EW}$.

\subsection{Limits from $\Sigma_u^u(\tst_{1,2})$}

Usually, the dominant contributions to $\Sigma_u^u$ come from
the top squarks, owing to the large value of the top-squark Yukawa coupling $f_t$.
The top squark contributions are given by\cite{ltr,rns}
\be
\Sigma_u^u (\tst_{1,2})= \frac{3}{16\pi^2}F(m_{\tst_{1,2}}^2)
\left[ f_t^2-g_Z^2\mp \frac{f_t^2 A_t^2-8g_Z^2
(\frac{1}{4}-\frac{2}{3}x_W)\Delta_t}{m_{\tst_2}^2-m_{\tst_1}^2}\right]
\label{eq:sigtuu}
\ee
where 
$\Delta_t=(m_{\tst_L}^2-m_{\tst_R}^2)/2+M_Z^2\cos 2\beta (\frac{1}{4}-
\frac{2}{3}x_W)$, $x_W\equiv\sin^2\theta_W$ and where
\be
F(m^2)= m^2\left(\log\frac{m^2}{Q^2}-1\right) .
\label{eq:F}
\ee
with the optimized scale choice $Q^2 =m_{\tst_1}m_{\tst_2}$.
In the denominator of Eq. \ref{eq:sigtuu},
the tree level expressions for $m_{\tst_{1,2}}^2$ should be used.

\begin{figure}[tbp]
\includegraphics[height=0.35\textheight]{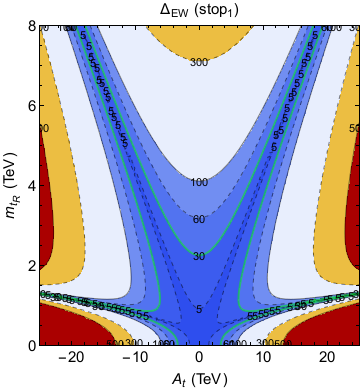}
\includegraphics[height=0.35\textheight]{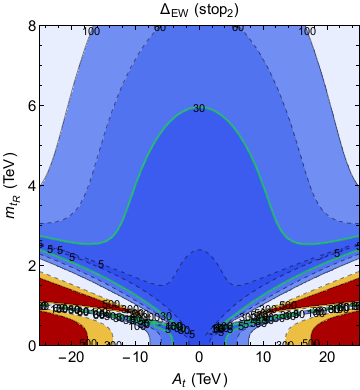}
\caption{Contour plot of {\it a}) $\Delta_{EW}(\tst_1 )$ from $\Sigma_u^u(\tst_1 )$
and {\it b}) $\Delta_{EW}(\tst_2 )$ from $\Sigma_u^u(\tst_2 )$ in the 
weak-scale $m_{\tst_R}$ vs. $A_t$ plane. 
The curves correspond to fixed values of 
$\Delta_{EW}(\tst_{1,2})$ as labelled.
\label{fig:stop}}
\end{figure}

In Fig. \ref{fig:stop}, we plot out the $\Sigma_u^u(\tst_{1,2})$ 
contributions to $\Delta_{EW}$ in the weak scale $m_{\tst_R}$ vs. $A_t$ plane
where we take $m_{\tst_L}= 2.6 m_{\tst_R}$ which is typical for RNS models. 
We also adopt $\tan\beta=10$, $\mu =150$ GeV and $f_t=0.8365$.
In frame {\it a}), the blue shaded region bounded by the green contour
has $\Delta_{EW}(\tst_1)<30$. For $A_t\sim 0$, then
$m_{\tst_R}\alt 2$ TeV although for this value of $A_t$ it is essentially impossible
to generate $m_h$ as high as 125 GeV\cite{h125}. 
For large stop mixing, {\it i.e.} large $|A_t|$, 
then there still exist bands of low $\Delta_{EW}(\tst_1 )$
at large $|A_t|$ and $m_{\tst_R}\sim 1$ TeV which occur where the mass
eigenvalue $m_{\tst_1}$ becomes very small, $\sim 100$ GeV, so that
$F(m_{\tst_1}^2)$ is suppressed. Alternatively, the low
$\Delta_{EW}(\tst_1 )$ bands at large $m_{\tst_R}$ and large $A_t$ occur
due to cancellations in the square bracket of Eq. \ref{eq:sigtuu}.

In frame {\it b}), where $\Delta_{EW}(\tst_2 )$ is shown, we again
see large regions of suppressed fine-tuning contributions. 
Here, $m_{\tst_R}\alt 6$ TeV is required for $\Delta_{EW}(\tst_2 )<30$
although side-bands extend out to large $|A_t|$ at $m_{\tst_R}\sim 1$ 
and 2.5 TeV. The upper band occurs where $m_{\tst_2}/m_{\tst_1}\sim e$ so 
that the log term in $F(m_{\tst_2}^2)$ cancels. The main point is that
regions exist with large enough $|A_t|$ (so that  $m_h\to 125$ GeV)
and $m_{\tst_R}\sim 1-3$ TeV where both $\Delta_{EW}(\tst_{1,2})$
become small\cite{ltr,rns}. For higher values of $m_{\tst_R}$, 
one or the other of $\Delta_{EW}(\tst_{1,2})$ necessarily becomes $>30$, 
leading to fine-tuning. 

\begin{figure}[tbp]
\includegraphics[height=0.4\textheight]{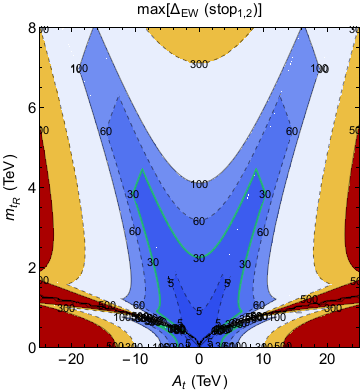}
\caption{Contour plot of $max [ \Delta_{EW}(\tst_1 ),\Delta_{EW}(\tst_2 ) ]$ 
in the weak-scale $m_{\tst_R}$ vs. $A_t$ plane.
The region below the thick dashed black contour is where $m_{\tst_1}<m_t$.
\label{fig:t1t2}}
\end{figure}

Since both $\Sigma(\tst_1)$ and $\Sigma (\tst_2)$ must be small, we show
in Fig. \ref{fig:t1t2} the resulting green contour from requiring 
$max[\Delta_{EW}(\tst_1),\Delta_{EW}(\tst_2)]<30$. These combined results 
show that $m_{\tst_R}$ is bounded from above by about 4.5 TeV. 
The region below the thick black dashed contour is where $m_{\tst_1}<m_t$.

\subsection{Limits from $\Sigma_u^u(\tb_{1,2})$}

\begin{figure}[tbp]
\includegraphics[height=0.35\textheight]{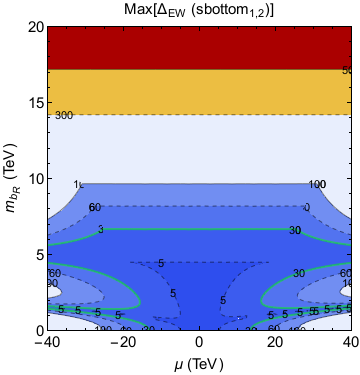}
\caption{Contour plot of $\Delta_{EW}(\tb_1,\tb_2 )$ 
from $\Sigma_u^u(\tb_{1,2} )$ in the 
weak-scale $m_{\tb_R}$ vs. $\mu$ plane.
We take $m_{\tb_L}=0.72 m_{\tb_R}$ and $\tan\beta =10$.
\label{fig:sbottom}}
\end{figure}

In Fig. \ref{fig:sbottom}, we plot the contributions to
$max[\Delta_{EW}(\tb_1),\Delta_{EW}(\tb_2)]$ from the $\tb_{1,2}$ squarks.
We plot for $\tan\beta =10$ and $m_{\tb_L}=0.72 m_{\tb_R}$ but in the $m_{\tb_R}$ vs. $\mu$ plane.
For $b$-squark contributions, we have
\be
\Sigma_u^u (\tb_{1,2})= \frac{3}{16\pi^2}F(m_{\tb_{1,2}}^2)
\left[ g_Z^2\mp \frac{f_b^2\mu^2-8g_Z^2
(\frac{1}{4}-\frac{1}{3}x_W)\Delta_b}{m_{\tb_2}^2-m_{\tb_1}^2}\right]
\label{eq:sigbuu}
\ee
where $\Delta_b=(m_{\tb_L}^2-m_{\tb_R}^2)/2-M_Z^2\cos 2\beta 
(\frac{1}{4}-\frac{1}{3}x_W)$. We take $f_b=0.13$ and $Q^2=m_{\tst_1}m_{\tst_2}$ 
with $m_{\tst_1}=1275$ GeV and $m_{\tst_2}=3690$ GeV.

From Fig. \ref{fig:sbottom}, the contribution $\Delta (\tb_1)$ can become small even for 
very large $m_{\tb_R}\sim 20$ TeV, but only for large/fine-tuned values of $\mu$. 
In contrast, $\Delta_{EW}(\tb_2) <30$
only for $m_{\tb_R}\alt 7$ TeV.
The latter provides a solid upper limit on $m_{\tb_2}$ which is usually $\sim m_{\tb_R}$. 

\subsection{Limits from $\Sigma_u^u(\tw_{1,2})$}

\begin{figure}[tbp]
\includegraphics[height=0.35\textheight]{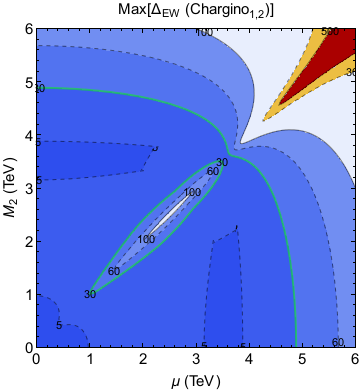}
\caption{Contour plot of $\Delta_{EW}(\tw_1,\tw_2 )$ from $\Sigma_u^u(\tw_1,\tw_2 )$ in 
the weak-scale $M_2$ vs. $\mu$ plane for $\tan\beta =10$.
\label{fig:Cinos}}
\end{figure}

The chargino contributions to $\Delta_{EW}$ are given by
\be
\Sigma_u^u(\tw_{1,2}^\pm )= \frac{-g^2}{16\pi^2} F(m_{\tw_{1,2}}^2)
\left(1\mp \frac{M_2^2+\mu^2-2m_W^2\cos 2\beta}{m_{\tw_2}^2-m_{\tw_1}^2}
\right) .
\label{eq:Sigw12}
\ee
We plot the $\tw_1$ and $\tw_2$ contributions to $\Delta_{EW}$ in
Fig's \ref{fig:Cinos}{\it a}), {\it b}) in the $M_2$ vs. $\mu$ plane.
The contribution $\Delta_{EW}(\tw_1 )$ is always small for $\mu \alt 1$ TeV. 
In fact, the $\Delta_{EW}(\tw_1 )$ contribution is small all over the plane
{\it except} when $\mu\simeq M_2$ in which case the denominator
in Eq. \ref{eq:Sigw12} becomes small so that $\Delta_{EW}(\tw_1 )$ blows up. 
In contrast, from $\Delta_{EW}(\tw_2 )$
we find that $\mu, \ M_2\alt 5$ TeV. This illustrates that wino masses 
all by themselves cannot become too large.

\subsection{Limits from $\Sigma_u^u(\tz_{1-4})$}

\begin{figure}[tbp]
\includegraphics[height=0.35\textheight]{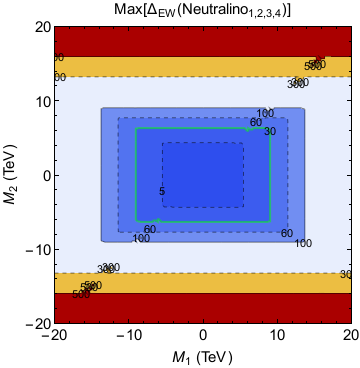}
\caption{Contour plots of $\Delta_{EW}(\tz_1,\tz_2,\tz_3,\tz_4)$ 
from $\Sigma_u^u(\tz_i )$ in the 
weak-scale $M_2$ vs. $M_1$ plane for $\mu =150$ GeV and $\tan\beta =10$.
\label{fig:Ninos}}
\end{figure}

The contributions to $\Delta_{EW}$ from the neutralino mass squared matrix
are found to be\cite{rns}:

\be
 \Sigma_u^u(\tz_i)= \frac{1}{16\pi^2}
  \frac{F(m_{\tz_i}^2)}{D(\tz_i)} \left[K(\tz_i)- 2(g^2+g'^2)\mu^2
  M_Z^2\cos^2\beta (m_{\tz_i}^2-m_{\pino}^2) \right] \, 
\ee
where 
\bea K(\tz_i)&=&
  -m_{\tz_i}^6 (g^2+g'^2) \nonumber\\ && +m_{\tz_i}^4 \left[ g^2
  (M_1^2+\mu^2)+g'^2 (M_2^2+\mu^2)+(g^2+g'^2)M_Z^2\right] \nonumber\\
  &&-m_{\tz_i}^2 \left[\mu^2 (g^2 M_1^2+g'^2 M_2^2)+(g^2+g'^2)M_Z^2
  m_{\pino}^2 \right], 
\eea
$D(\tz_i)=\prod_{j\neq i} (m^2_{\tz_i}-m^2_{\tz_j})$ 
and $m_{\pino} = M_1 \cos^2\theta_W +M_2\sin^2\theta_W$.
The contribution to $max[\Delta_{EW}(\tz_i )]$ ($i=1-4$) 
are shown in Fig. \ref{fig:Ninos} in the $M_1$ vs. $M_2$ plane for 
$\mu =150$ GeV and $\tan\beta =10$.
From the figure we are able to extract that $M_1\alt 10$ TeV
while $M_2\alt 6$ TeV.
These bounds are independent of any assumptions about gaugino 
mass unification, unlike those of Sec. \ref{sec:mass}.

\subsection{Limits from $\Sigma_u^u(h,H,H^\pm )$}
\begin{figure}[tbp]
\includegraphics[height=0.35\textheight]{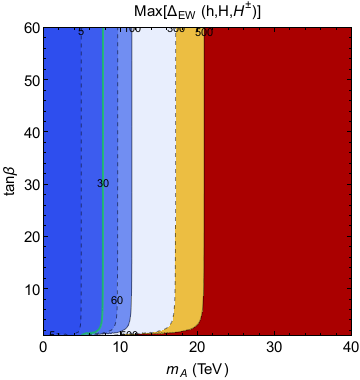}
\caption{Contour plot of $max[\Delta_{EW}(h,H,H^\pm )]$ 
from $\Sigma_u^u(h,H,H^\pm )$ in the 
weak-scale $\tan\beta$ vs. $m_A$ plane.
\label{fig:H}}
\end{figure}

For Higgs bosons, it is found that\cite{rns}
\bea
\Sigma_u^u(h,H)&=& \frac{g_Z^2}{16\pi^2} F(m_{h,H}^2)\left(
1\mp \frac{M_Z^2+m_A^2 (1+4\cos 2\beta +2\cos^2 2\beta )}{m_{H}^2-m_{h}^2}
\right)
\eea
while 

\bea
\Sigma_u^u(H^\pm )=\frac{g^2}{32\pi^2} F(m_{H^\pm}^2) .
\eea
The contributions from each of $h$, $H$ and $H^\pm$ are plotted in 
Fig. \ref{fig:H} as $max[\Delta_{EW}(h),\Delta_{EW}(H),\Delta_{EW}(H^\pm)]$ 
in the $m_A$ vs. $\tan\beta$ plane.
The contribution $\Delta_{EW}(h,H,H^\pm )<30$ requires $m_A< 8$ TeV.
The largest term usually comes from the charged Higgs contribution.
These values are somewhat higher than that which comes from the $m_{H_d}^2$ term in Eq. \ref{eq:mzs}.
%

%

%
\end{document}